\documentclass[onefignum,onetabnum]{siamart171218}

\usepackage{lipsum}
\usepackage{amsfonts}
\usepackage{graphicx}
\usepackage{epstopdf}
\usepackage{algorithmic}
\ifpdf
  \DeclareGraphicsExtensions{.eps,.pdf,.png,.jpg}
\else
  \DeclareGraphicsExtensions{.eps}
\fi

\newsiamremark{remark}{Remark}
\newsiamremark{hypothesis}{Hypothesis}
\crefname{hypothesis}{Hypothesis}{Hypotheses}
\newsiamthm{claim}{Claim}

\headers{Tridiagonal Factorization of Skew-Symmetric Matrices}{Ishna Satyarth, Chao Yin, et al.}

\title{Performant Tridiagonal  Factorization of Skew-Symmetric Matrices\\[6pt]
FAMLIES Working Note \#1\thanks{Submitted to the editors DATE.
\funding{This work was funded at SMU in part by the National Science Foundation (grants CHE-2143725 and OAC-2003931) and the US Department of Energy (grants DE-SC0022893 and DE-SC0025950), and at UT by the National Science Foundation (grants OAC-2003921 and CCF-2446145) and a gift from Arm.}}}

\author{
Ishna Satyarth{\thanks{\MakeLowercase{\uppercase{D}epartment of \uppercase{C}omputer \uppercase{S}cience, \uppercase{S}outhern \uppercase{M}ethodist \uppercase{U}niversity, \uppercase{D}allas, \uppercase{TX}
  (\email{isatyarth@smu.edu}).}}}\,\,\thanks{These authors contributed equally.}
\and 
Chao Yin\footnotemark[4]\,\,\footnotemark[3]
\and 
Devin A. Matthews\thanks{Department of Chemistry, Southern Methodist University, Dallas, TX
  (\email{chaoy@smu.edu}, \email{damatthews@smu.edu}).}
  \and 
Maggie Myers\thanks{Oden Institute, The University of Texas at Austin, Austin, TX
  (\email{myers@cs.utexas.edu}, \email{rvdg@cs.utexas.edu}).}
  \and 
Robert van de Geijn\footnotemark[5]
  \and 
Ruqing G. Xu\thanks{NVIDIA, Santa Clara, CA
  (\email{ruqingx@nvidia.com}).}
  }
  
\usepackage{amsopn}

\makeatletter
\newcommand*{\addFileDependency}[1]{
  \typeout{(#1)}
  \@addtofilelist{#1}
  \IfFileExists{#1}{}{\typeout{No file #1.}}
}
\makeatother

\colorlet{review}{black}
\colorlet{review2}{black}

\renewcommand\arraystretch{1.3}
\ifpdf
\hypersetup{
  pdftitle={Performant Tridiagonal Matrix Factorizations of Skew-Symmetric Matrices},
  pdfauthor={I. Satyarth, C. Yin, D.A. Matthews, M. Myers, R. van de Geijn, and R. Xu}
}
\fi

\usepackage{array}
\usepackage{booktabs}
\usepackage{tikz}
\usepackage{multirow}
\usepackage{colortbl}
\usepackage{ifthen}
\usepackage{wrapfig}

\newcommand{\LTLt}{{\sc LTLt}}
\newcommand{\skewLTLt}{{\sc skewLTLt}}

\DeclareRobustCommand\sampleline[1]{%
\raisebox{-0.55em}{
  \tikz\draw[#1] (-0.8em,0)
  -- (0.8em,0)
  (0,0.8em)
  -- (0,-0.8em);}%
}

\newcolumntype{I}{!{\vrule width 1.5pt}}
\newlength\savedwidth
\newcommand\whline{\noalign{\global\savedwidth\arrayrulewidth
                            \global\arrayrulewidth 1.05pt}%
           \hline
           \noalign{\global\arrayrulewidth\savedwidth}}

\newcommand{\FlaTwoByTwo}[4]{
\left(
\begin{array}{c I c}
#1 & #2 \\ \whline
#3 & #4
\end{array}
\right)
}

\newcommand{\FlaThreeByThreeBR}[9]{
\left(
\begin{array}{c I c | c}
#1 & #2 & #3 \\ \whline 
#4 & #5 & #6 \\ \hline
#7 & #8 & #9 
\end{array}
\right) 
}

\newcommand{\operation}{}

\newcommand{\routinename}{}

\newcommand{\precondition}{~}

\newcommand{\postcondition}{~}

\newcommand{\invariant}{~}

\newcommand{\guard}{~}

\newcommand{\partitionings}{~}

\newcommand{\partitionsizes}{~}

\newcommand{\blocksize}{blank}

\newcommand{\repartitionings}{~}

\newcommand{\repartitionsizes}{~}

\newcommand{\moveboundaries}{~}

\newcommand{\beforeupdate}{~}

\newcommand{\afterupdate}{~}

\newcommand{\update}{~}

\newcommand{\resetsteps}{

\renewcommand{\operation}{\phantom{[A] = op( A )}}

\renewcommand{\routinename}{\operation}

\renewcommand{\precondition}{\phantom{A = \widehat A}}

\renewcommand{\postcondition}{\phantom{A = \widehat A}}

\renewcommand{\invariant}{\phantom{ \FlaTwoByTwo{A_{TL}}{A_{TR}}{A_{BL}}{A_{BR}} =
		\FlaTwoByTwo{A_{TL}}{A_{TR}}{A_{BL}}{A_{BR}}
		\wedge
		\FlaTwoByTwo{A_{TL}}{A_{TR}}{A_{BL}}{A_{BR}} =
		\FlaTwoByTwo{A_{TL}}{A_{TR}}{A_{BL}}{A_{BR}}~~~~~~~
		}}

\renewcommand{\blocksize}{blank}

\renewcommand{\guard}{\phantom{m( A_{BL} ) < m( A )}}

\renewcommand{\partitionings}{
$
\phantom{\FlaTwoByTwo{A_{TL}}{A_{TR}}{A_{BL}}{A_{BR}}
\rightarrow
\FlaThreeByThreeBR
   {A_{00}}{a_{01}}{A_{02}}
   {a_{10}^T}{\alpha_{11}}{a_{12}^T}
   {A_{20}}{a_{21}}{A_{22}}}   
$
}

\renewcommand{\partitionsizes}{$ \phantom{m( A )} $}

\renewcommand{\repartitionings}{
$
\phantom{\FlaTwoByTwo{A_{TL}}{A_{TR}}{A_{BL}}{A_{BR}}
\rightarrow
\FlaThreeByThreeBR
   {A_{00}}{a_{01}}{A_{02}}
   {a_{10}^T}{\alpha_{11}}{a_{12}^T}
   {A_{20}}{a_{21}}{A_{22}}}   
$
}

\renewcommand{\repartitionsizes}{$\phantom{m(A)}$}

\renewcommand{\moveboundaries}{
$
\phantom{\FlaTwoByTwo{A_{TL}}{A_{TR}}{A_{BL}}{A_{BR}}
\rightarrow
\FlaThreeByThreeBR
   {A_{00}}{a_{01}}{A_{02}}
   {a_{10}^T}{\alpha_{11}}{a_{12}^T}
   {A_{20}}{a_{21}}{A_{22}}}   
$
}

\renewcommand{\beforeupdate}{
\phantom{\FlaTwoByTwo{A_{TL}}{A_{TR}}{A_{BL}}{A_{BR}}
\rightarrow
\FlaThreeByThreeBR
   {A_{00}}{a_{01}}{A_{02}}
   {a_{10}^T}{\alpha_{11}}{a_{12}^T}
   {A_{20}}{a_{21}}{A_{22}}}   
}

\renewcommand{\afterupdate}{
\phantom{\FlaTwoByTwo{A_{TL}}{A_{TR}}{A_{BL}}{A_{BR}}
\rightarrow
\FlaThreeByThreeBR
   {A_{00}}{a_{01}}{A_{02}}
   {a_{10}^T}{\alpha_{11}}{a_{12}^T}
   {A_{20}}{a_{21}}{A_{22}}}   
}

\renewcommand{\update}{
\phantom{$
\begin{array}{l}
\\
\\
\\
\end{array}
$}
}
}

\newcommand{\NoShow}[1]{}

\newcommand{\moreinitialize}{}

\newcommand{\FlaAlgorithm}{
\begin{tabular}{|p{4.9in}|} \hline
$\mbox{\color{blue}Algorithm:~}\routinename$
\\ \hline
\partitionings \\
$\mbox{\color{blue} ~~~where~}$ \partitionsizes 
\moreinitialize 
\\ 
$\mbox{\color{blue}while~} \ShowGuard \mbox{~\color{blue} do}$
\\
\ifthenelse{\equal{\blocksize}{1}}{}%
{%
\ifthenelse{ \equal{\blocksize}{blank} }{}%
{~~~~{\bf Determine block size $ \blocksize $}\\}%
}
\repartitionings 
\NoShow{\\
~$\mbox{\color{blue} ~~~where~}$ \repartitionsizes
}
\\ \hline  \update 
\\ \hline
\moveboundaries 
\\
$\mbox{\color{blue} endwhile} $
\\ \hline 
\end{tabular}
}

\newcounter{WSStep}
\newcommand{\ShowPrecondition}{\ifthenelse{\value{WSStep}<1}%
   {{\color{white} \precondition}}
   {\ifthenelse{\value{WSStep}=1}%
    {\color{red} \precondition}
    {\color{black} \precondition}}}
\newcommand{\ShowPostcondition}{\ifthenelse{\value{WSStep}<1}%
   {{\color{white} \postcondition}}
   {\ifthenelse{\value{WSStep}=1}%
    {\color{red} \postcondition}
    {\color{black} \postcondition}}}
\newcommand{\ShowInvariant}{\ifthenelse{\value{WSStep}<2}%
   {{\color{white} \invariant}}
   {\ifthenelse{\value{WSStep}=2}%
    {\color{red} \invariant}
    {\color{black} \invariant}}}
\newcommand{\ShowGuard}{\ifthenelse{\value{WSStep}<3}%
   {{\color{lightgray!25} \guard}}
   {\ifthenelse{\value{WSStep}=3}%
    {\color{red} \guard}
    {\color{black} \guard}}}
\newcommand{\ShowGuardTwo}{\ifthenelse{\value{WSStep}<3}%
   {{\color{white} \guard}}
   {\ifthenelse{\value{WSStep}=3}%
    {\color{red} \guard}
    {\color{black} \guard}}}
\newcommand{\ShowPartitionings}{\ifthenelse{\value{WSStep}<4}%
   {{\color{lightgray!25} \partitionings}}%
   {\ifthenelse{\value{WSStep}=4}%
    {\color{red} \partitionings}%
    {\color{black} \partitionings}}}
\newcommand{\ShowPartitionSizes}{\ifthenelse{\value{WSStep}<4}%
   {{\color{lightgray!25} \partitionsizes}}
   {\ifthenelse{\value{WSStep}=4}%
    {\color{red} \partitionsizes}
    {\color{black} \partitionsizes}}}

\newcommand{\ShowRepartitionings}{\ifthenelse{\value{WSStep}<5}%
   {{\color{lightgray!25} \repartitionings}}
   {\ifthenelse{\value{WSStep}=5}%
    {\color{red} \repartitionings}
    {\color{black} \repartitionings}}}
    
\newcommand{\ShowRepartitionSizes}{\ifthenelse{\value{WSStep}<5}%
   {{\color{lightgray!25} \repartitionsizes}}
   {\ifthenelse{\value{WSStep}=5}%
    {\color{red} \repartitionsizes}
    {\color{black} \repartitionsizes}}}
    
\newcommand{\ShowMoveBoundaries}{\ifthenelse{\value{WSStep}<5}%
   {{\color{lightgray!25} \moveboundaries}}
   {\ifthenelse{\value{WSStep}=5}%
    {\color{red} \moveboundaries}
    {\color{black} \moveboundaries}}}

\newcommand{\ShowBeforeUpdate}{\ifthenelse{\value{WSStep}<6}%
   {{\color{white} \beforeupdate}}
   {\ifthenelse{\value{WSStep}=6}%
    {\color{red} \beforeupdate}
    {\color{black} \beforeupdate}}}
\newcommand{\ShowAfterUpdate}{\ifthenelse{\value{WSStep}<7}%
   {{\color{white} \afterupdate}}
   {\ifthenelse{\value{WSStep}=7}%
    {\color{red} \afterupdate}
    {\color{black} \afterupdate}}}
\newcommand{\ShowUpdate}{\ifthenelse{\value{WSStep}<8}%
   {{\color{lightgray!25} \update}}
   {\ifthenelse{\value{WSStep}=8}%
    {\color{red} \update}
    {\color{black} \update}}}

\newcommand{\FlaWorksheetNine}{
\begin{tabular}{| c | p{0.9\textwidth} |}\hline
{\color{white}Step} & $\mbox{\color{blue}Algorithm:~}\routinename$
\\ \hline
 &%
$ \phantom{\left\{ 
\begin{minipage}{0.88\textwidth} 
$\ShowPrecondition$  
\end{minipage}
\right\}}
$%
\\ \hline
\rowcolor{lightgray!25}   
& %
\begin{minipage}{0.88\textwidth}%
\vspace{0.05in}
\ShowPartitionings~ \\
\mbox{\color{blue} ~~~where~} \ShowPartitionSizes
\end{minipage}
\\ \hline
& 
$ \phantom{\left\{ 
\begin{minipage}{0.88\textwidth} 
$\ShowInvariant $
\end{minipage}
\right\}} $ 
\\ \hline
\rowcolor{lightgray!25}   
&$\mbox{\color{blue}while~} \ShowGuard \mbox{~\color{blue} do}$
\\ \hline 
 &  
$
\phantom{\left\{
\begin{minipage}[t]{0.88\textwidth}%
~~~~$
\ShowInvariant 
\wedge \ShowGuardTwo$
\end{minipage}
\right\}}
$ 
\\ \hline
\rowcolor{lightgray!25}   
 & ~~~~ \begin{minipage}{0.85\textwidth}%
\vspace{0.05in}
\ifthenelse{\equal{\blocksize}{1}}{}%
{%
\ifthenelse{ \equal{\blocksize}{blank} }{}%
{{\bf Determine block size $ \blocksize $}\\}%
}
\ShowRepartitionings~ \\
$\mbox{\color{blue} ~~~where~}$ \ShowRepartitionSizes
\end{minipage}
\\ \hline
& 
$ \phantom{\left\{ 
\begin{minipage}{0.88\textwidth} 
~~~~ \ShowBeforeUpdate 
\end{minipage}
\right\}}
$
\\ \hline
\rowcolor{lightgray!25}  
 & ~~~~  \ShowUpdate 
\\ \hline 
& 
$ \phantom{\left\{ 
\begin{minipage}{0.88\textwidth} 
~~~~ \ShowAfterUpdate 
\end{minipage}
\right\}}
$
\\ \hline
\rowcolor{lightgray!25}   
 & ~~~~ \begin{minipage}{0.85\textwidth}%
\vspace{0.05in}
\ShowMoveBoundaries~
\end{minipage}
\\ \hline
& 
$ \phantom{\left\{ 
\begin{minipage}{0.88\textwidth} 
~~~~ $ \ShowInvariant  $ 
\end{minipage}
\right\}}
$
\\ \hline
\rowcolor{lightgray!25}  
 &$\mbox{\color{blue} endwhile} $
\\ \hline 
& 
$ \phantom{\left\{ 
\begin{minipage}{0.88\textwidth} 
$ \ShowInvariant \wedge \neg( \ShowGuardTwo )$ 
\end{minipage}
\right\}}
$
\\ \hline
& 
$ \phantom{\left\{ 
\begin{minipage}{0.88\textwidth} 
$ \ShowPostcondition $ 
\end{minipage}
\right\}}
$
\\ \hline
\end{tabular}
}

\newcommand{\FlaCostWorksheet}{
\begin{tabular}{| c | p{0.45\textwidth}
p{0.45\textwidth}|}\hline
Step & $\mbox{\color{blue}Algorithm:~}\routinename $ &
\\ \hline
1a &%
$ \left\{ 
\begin{minipage}{0.44\textwidth} 
$\ShowPrecondition$  
\end{minipage}
\right\}
$
&
\\ \hline
\rowcolor{lightgray!25}   
4 & %
\begin{minipage}{0.88\textwidth}%
\vspace{0.05in}
\ShowPartitionings~ \\
\mbox{\color{blue} ~~~where~} \ShowPartitionSizes
\end{minipage}
& 
\begin{minipage}{0.44\textwidth}
\hfill \CostInit
\end{minipage}
\\ \hline
2 & 
$ \left\{ 
\begin{minipage}{0.44\textwidth} 
$\ShowInvariant $
\end{minipage}
\right\} $ 
&
\begin{minipage}{0.44\textwidth}
$ \{ $ \hfill \CostInvariant
$ \} $ 
\end{minipage}
\\ \hline
\rowcolor{lightgray!25}   
3 &$\mbox{\color{blue}while~} \ShowGuard \mbox{~\color{blue} do}$
&
\\ \hline 
2,3 &  
$
\left\{
\begin{minipage}[t]{0.41\textwidth}%
$
~~~~ \ShowInvariant 
\wedge \ShowGuardTwo$
\end{minipage}
\right\}
$ 
&
\begin{minipage}{0.44\textwidth}
$ \{ $ \hfill \CostInvariant
$ \} $
\end{minipage}
\\ \hline
\rowcolor{lightgray!25}   
5a & ~~~~ \begin{minipage}{0.41\textwidth}%
\vspace{0.05in}
\ifthenelse{\equal{\blocksize}{1}}{}%
{%
\ifthenelse{ \equal{\blocksize}{blank} }{}%
{{\bf Determine block size $ \blocksize $}\\}%
}
\ShowRepartitionings~ 
\NoShow{\\
$\mbox{\color{blue} ~~~where~}$ \ShowRepartitionSizes}
\end{minipage}
&
\\ \hline
6 & 
$ \left\{ 
\begin{minipage}{0.41\textwidth} 
~~~~ \ShowBeforeUpdate 
\end{minipage}
\right\}
$
&
$ \{ $ \hfill 
\CostBefore
$ \} $
\\ \hline
\rowcolor{lightgray!25}  
8 & 
\begin{minipage}{0.41\textwidth} 
~~~~  \ShowUpdate 
\end{minipage}
&
\hfill 
$
C := C + 2 
$
\\ \hline 
7 & 
$ \left\{ 
\begin{minipage}{0.41\textwidth} 
~~~~ \ShowAfterUpdate 
\end{minipage}
\right\}
$
&
$ \{ $ \hfill 
\CostAfter
$ \} $
\\ \hline
\rowcolor{lightgray!25}   
5b & ~~~~ \begin{minipage}{0.41\textwidth}%
\vspace{0.05in}
\ShowMoveBoundaries~
\end{minipage}
&
\\ \hline
2 & 
$ \left\{ 
\begin{minipage}{0.44\textwidth} 
~~~~ $ \ShowInvariant  $ 
\end{minipage}
\right\}
$
&
\begin{minipage}{0.44\textwidth}
$ \{ $ \hfill \CostInvariant
$ \} $
\end{minipage}
\\ \hline
\rowcolor{lightgray!25}  
 &$\mbox{\color{blue} endwhile} $
 &
\\ \hline 
2,3 & 
$ \left\{ 
\begin{minipage}{0.44\textwidth} 
$ \ShowInvariant \wedge \neg( \ShowGuardTwo )$ 
\end{minipage}
\right\}
$
&
\begin{minipage}{0.44\textwidth}
$ \{ $ \hfill \CostInvariant
$ \} $
\end{minipage}
\\ \hline
1b & 
$ \left\{ 
\begin{minipage}{0.44\textwidth} 
$ \ShowPostcondition $ 
\end{minipage}
\right\}
$
&
\begin{minipage}{0.44\textwidth}
$ \{ $ \hfill \CostPostCond
$ \} $
\end{minipage}
\\ \hline
\end{tabular}
}

\usepackage{listings}

\begin{document}

\maketitle

\begin{abstract}
  The factorization of skew-symmetric matrices is a critically understudied area of dense linear algebra, particularly in comparison to that of general and symmetric matrices. While some algorithms can be adapted from the symmetric case, the cost of algorithms can be reduced by exploiting skew-symmetry. This work examines the factorization of a skew-symmetric matrix $ X $ into its $ L T L^T $ decomposition, where $ L $ is unit lower triangular and $ T $ is tridiagonal.  This is also known as a triangular tridiagonalization.
This operation is a means  for computing the determinant of $X$ as the square of the (cheaply-computed) Pfaffian of the skew-symmetric tridiagonal matrix $T$ as well as for solving systems of equations, across fields such as quantum electronic structure and machine learning. Its application also often requires pivoting in order to improve numerical stability. 
We compare and contrast  previously-published algorithms with those systematically derived using the FLAME methodology. Performant parallel CPU implementations are achieved by fusing operations  at multiple levels in order to reduce memory traffic overhead.
A key factor is the employment of new capabilities of the  BLAS-like Library Instantion Software (BLIS) framework, which now supports casting level-2 and level-3 BLAS-like operations by leveraging its {\sc gemm} and other kernels, hierarchical parallelism, and cache blocking. 
 A prototype,  concise C++ API  facilitates the translation of correct-by-construction algorithms into correct code.
Experiments verify that  the resulting implementations greatly exceed the performance of previous work.

\end{abstract}

\begin{keywords}
  skew-symmetric matrices, linear algebra, matrix factorization, high-performance computing, formal methods, shared-memory parallelization
\end{keywords}

\begin{AMS}
  68Q25, 68W10, 81V74, 15B57, 15A23, 15A15
\end{AMS}

\newcommand{\comment}[1]{}
\section{Introduction}

Under well-understood conditions, a skew-symmetric indefinite matrix $ X $ can be factored as $ P X P^T = L T L^T $, where $ P $ is a permutation matrix, $ L $ is a unit lower-triangular matrix and $ T $ is a skew-symmetric tridiagonal matrix.
This is referred to as {\em triangular tridiagonalization}~\cite{Miroslav2011}.
It is a variation on the Cholesky%
\NoShow{($ X = L L^T $)} and $ L D L^T $ (where $ D $ is diagonal) factorizations for symmetric matrices.
We are motivated by the computation of the Pfaffian $ {\rm Pf}(X) $, defined as $ {\rm Pf}( X ) = \frac{1}{ 2^n n! } \sum_{ \sigma \in S_{ 2n } } {\rm sgn}( \sigma ) \prod_i^n x_{ \sigma( 2i-1 ), \sigma( 2i ) } $  for skew-symmetric $ X $ of size $ 2n \times 2n $.  Here, $ S_{ 2n } $ represents the $ 2n $-element permutation set. 
It can be shown that $ {\rm Pf}( X )^2 = \det( X ) $. 
Also, if $ P X P^T = L T L^T $, where 
{
\renewcommand{\arraystretch}{1.0}
\footnotesize
\[
\label{eqn:T}
T = \left( \begin{array}{c c c c c}
0 & -\tau_{1,0} & 0 & \cdots & 0 \\
\tau_{1,0} & 0 & - \tau_{2,1} & \cdots & 0 \\
0 & \tau_{2,1} & 0 & \ddots & 0 \\
\vdots & \vdots & \ddots & \ddots & \vdots \\
0 & 0 & 0 & \cdots & 0
\end{array}
\right),
\]
}%
then $ {\rm Pf}( X ) = {\rm Pf}( T ) = \tau_{1,0}
\times \tau_{3,2}  \times \cdots
\times \tau_{2n-1,2n-2} $.
This quantity arises frequently in physics studies where pairs of Fermions are involved, such as the 2-dimensional Ising spin glass~\cite{thomas2009} and electronic structure quantum Monte Carlo~\cite{bajdich2009}. \textcolor{review}{For example, Variational Monte Carlo (VMC) with geminal wavefunctions requires rapid sampling of the wavefunction through the Pfaffian of the skew-symmetric matrix of geminal coefficients \cite{blockedvmc}. In machine learning, the regularized antisymmetric weight matrix $W-W^T-\gamma I$ appears in modeling of non-diffusive processes using deep graph networks \cite{gravina2023antisymmetricdgnstablearchitecture,10.1007/978-3-031-74643-7_3} and graph convolutional networks \cite{10.1093/bioinformatics/btae603}. 
\color{review} More generally, the skew-symmetric $LTL^T$ factorization fills a similar role for skew-symmetric matrix as does the related Cholesky ($LL^T$), $LDL^T$, and $LU$ factorizations for other matrix structures and completes this family of fundamental matrix decompositions.}
The $ L T L^T $ factorization can also be used as a step towards  solving the linear system $ X v = w $.

Classic work focused on computing an $ L T L^T $ factorization starting with a {\em symmetric} matrix $ X$~\cite{Aasen,ParlettReid}.  A high-performing blocked algorithm for this was proposed by Rozlo\v{z}n\'{\i}k et al.~\cite{Miroslav2011}.  Wimmer~\cite{Wimmer2012} adapted these for the skew-symmetric case and proposed novel two-step algorithms.  Improvements to Wimmer's work were recently reported by Xu et al~\cite{blockedvmc}.

More than two decades ago, the still-evolving FLAME methodology was introduced for systematically deriving algorithms hand in hand with their proofs of correctness~\cite{FLAME}.
This approach has been applied to a broad class of dense linear algebra operations as well as Krylov subspace methods~\cite{Eijkhout20101805} and graph algorithms~\cite{TC_Correctness,9835383}.  
A chapter on this, and its impact, can be found in a book dedicated to Edsger Dijkstra~\cite{10.1145/3544585,10.1145/3544585.3544597}, who inspired the approach.  
In a companion technical report~\cite{vandegeijn2023derivingalgorithmstriangulartridiagonalization}, we apply this methodology to the $ L T L^T $ factorization of a skew-symmetric matrix, allowing us to repeat only brief highlights of the derivation process here while still making this paper self-contained.
Importantly, the FLAME approach yields a family of algorithms from which the ones most appropriate (typically, highest performing) 
can be chosen.

The present paper makes a number of contributions:
\begin{itemize}
\item 
Our best implementations of skew-symmetric triangular tridiagonalization outperform  those  in the only 
software  that we  are  aware  of,  PFAPACK~\cite{Wimmer2012,pfapack} and Pfaffine~\cite{blockedvmc,Pfaffine}, as reported  in Figure~\ref{fig:comparison}d, while also obtaining similar or better performance than software for related symmetric matrix factorizations.
\item 
It extends the scope of the FLAME methodology~\cite{Bientinesi:2008:FAR,FLAME,FLAME:API,Recipe,Quintana-Orti:2003:FDA,10.1145/3544585.3544597}.
    \item
    \textcolor{review}{It 
comprehensively
    compares and contrasts known and novel algorithms using FLAME-like notation~\cite{FLAME}.}
    \item 
    It contributes new algorithms (without and with pivoting), 
    including three blocked right-looking algorithms and one left-looking algorithm, some of\break{}which 
    attain higher performance and require less workspace.
\item 
   It exposes how more \textcolor{review}{straightforward} and efficient blocked right-looking algorithms are related to 
   Wimmer's two-step unblocked algorithm.
    \item 
    It identifies new level-2 and level-3 BLAS-like operations 
    for these algorithms.
    \NoShow{including 
a ``sandwiched'' (skew-)symmetric rank-k update: $ A T A^T $ and a ``sandwiched'' general matrix-matrix multiplication: $ A T B $.
   Here $ A$ and $ B$  are general matrices  and $ T $  is a skew-symmetric tridiagonal matrix.}
   \item 
   It shows that operation fusion and improved use of 
   bandwidth via parallelization of level-2 BLAS-like operations are essential for 
   high performance.
   \item 
   It discusses optimizing new BLAS-like operations via mechanisms recently added to the BLAS-like Library Instantiation Software (BLIS)~\cite{BLIS1,BLIS2,BLIS20}.
   
    \item 
    It lays the groundwork for a future C++ FLAME API for translating  algorithms presented in FLAME notation into code. This API  accommodates the  additional partitioning of matrices and vectors encountered for this operation  \textcolor{review}{and enables high performance for both blocked and unblocked algorithms.}

\end{itemize}


 \NoShow{
{\bf Begin of RuQing's sketch of an  intro}

[Perhaps begin with a comment on the LAPACK standard, how important / influencing it is, and in what aspects it's not perfect.]
One aspect that LAPACK didn't manage to cover is the various transformations of skew-symmetric matrices.
[Some comments here. Maybe borrow a few ideas from Wimmer's paper.]
The $LTL^T$ factorization is one of the important problems [Cite:LTLt] comparable to the Cholesky factorization of a symmetric or Hermitian matrix.
From $LTL^T$'s affinity to Cholesky factorization, traditional wisdom suggests implementing via Gaussian elimination processing from top-right to bottom-left (or to say: eliminating elements s.t. the bottom-right leftover becomes smaller and smaller?), from which comes the Parlett-Reid algorithm [Cite:Parlett-Reid] whose computational hotspot lies in the skew-symmetric rank-2 update \textsc{skr2}: $X = X + ab^T - ba^T$. The Parlett-Reid algorithm is extended into its first block-update formation in [Cite:Wimmer] to speed up its backend performance where \textsc{skr2} calls got grouped into \textsc{skr2k} for rank-$2k$ updates: $X = X + AB^T - BA^T$. Various efforts followed to improve further the efficiency of the \textsc{skr2k} backend [Cite:Gihō,RX]. However, the Parlett-Reid algorithm features a computational cost of $\frac23 m^3$ flops, doubling that of a Cholesky factorization. From intuition, a skew-symmetric matrix has around the same (if not fewer) degrees of freedom as a symmetric one ($\frac12 m(m-1)$ vs $\frac12 m(m+1)$. TODO: rewrite this sentence), hence their factorizations should come at the same cost at optimal conditions.
\NoShow{
Further, [Cite:Wimmer] indicates that $X$'s determinant can be obtained via $\det X = \left(\mathrm{Pf} X\right)^2$ where $\mathrm{Pf} X$ is a quantity called \emph{Pfaffian} [Cite:Pfaffian] and is computable by only factoring odd columns of $X$:
\[
   X = L_P T_P L_P^T \text{ where } T_P = \left(
   \begin{array}{cccccc}
        0      &-a_1   & 0      & 0      & \cdots & 0 \\
        a_1    & 0     &-b_{32} &-b_{42} & \cdots &-b_{m2} \\
        0      & b_{32}& 0      &-a_3    & \cdots & 0 \\
        0      & b_{42}& a_3    & 0      & \cdots & -b_{m4} \\
        \vdots & \vdots& \vdots & \vdots & \ddots & 0 \\
        0      & b_{m2}& 0      & b_{m4} & \cdots & 0
   \end{array}
   \right),
\]
which again demands only $\frac13 m^3$ flops, contrasting the regular perception that the determinant requires a full factorization to determine.
}
The most apparent difference that causes this inconsistency is that the Cholesky factorization does a rank-1 update for each row and column eliminated, but Parlett-Reid demands a rank-2 update \emph{to keep $X$ in its form}. This prompts us to deviate our algorithms' layout away from their traditional form so that this symmetry-caused computational ``redundancy'' is obviated.

In this paper, we present an approach based on the FLAME [Cite:FLAME] notation that systematically derives the algorithm for this skew-symmetric $LTL^T$ factorization through a post-condition-based methodology. We discovered that the algorithmic variant that reflects already-computed parts of $L$ (which we will later call a \emph{left-looking} variant) halves the cost of the original algorithm which reflects changes on the not-yet-computed parts of $X$ (later called \emph{right-looking} variant). More interestingly, if we approach the blocked variant of the original algorithm through this systematic post-condition-based methodology, allowing update blocks to break that formal symmetry of \textsc{skr2k}, it is again possible to reduce the computational cost from $\frac23 m^3$ flops to $\frac13 m^3$ flops.
Discovery of this new algorithm manifests the FLAME notation's approach by showing that the systematic derivation of an algorithm can effectively spot the path toward the post-conditions without extra redundancy the traditional constructions might have. It also populates the set of skew-symmetric linear algebra routines that LAPACK did not manage to cover.
} 

\section{Preliminaries}

We relate
skew-symmetric matrices and  Gauss transforms \textcolor{review}{to their} use in algorithms for computing the $ L T L^T $ decomposition.

\subsection{Notation}

\NoShow{
\begin{figure}
\begin{tabular}{ >{\centering}m{0.15\columnwidth} m{0.75\columnwidth} }
\toprule
$A$ & Upper-case Roman letters are used for matrices  \\
$a$ & Lower-case Roman letters are used for (column) vectors  \\
$\alpha$ & Lower-case Greek letters are used for real-valued scalars  \\
$e_f$ & Standard basis vector with a $1$ in the \underline{$f$}irst position \\
$e_l$ & Standard basis vector with a $1$ in the \underline{$l$}ast position \\
\textcolor{blue}{${A}$, ${a}$, ${\alpha}$} & Blue identifies sub-partitions that have already been computed at the current step \\
$\Big(\!\sampleline{black}\Big)$ & Partitioned matrix---the size of each sub-partition is implicit \\
$\Big(\!\sampleline{black,line width=1.5pt}\Big)$ & Partitioned matrix---a thick line typically separates regions of the matrix according to the current progress of a loop-based algorithm
\\
$TL, TM, \ldots $ &
Identify 
Top-Left, Top-Middle, etc. subparts of matrices
\\
$ \star $ &
Implicit (skew-)symmetric part of matrix, assuming only the lower triangular part is stored \\
\bottomrule
\end{tabular}
\caption{A summary of the notational conventions used in this work. The symbols $A$, $a$, and $\alpha$ are used to denote arbitrary matrices, vectors, and scalars.}
\label{table:notation}
\end{figure}
}

We adopt {\em Householder notation} where, as a general rule, matrices, (column) vectors, and scalars are denoted with upper-case Roman, lower-case Roman, and lower-case Greek letters, respectively.
Indexing starts at $ 0 $.
The $ m \times m $ identity matrix, $ I $, is partitioned by columns into standard basis vectors $ e_i $ as $
\renewcommand{\arraystretch}{1.0} I = 
\left( \begin{array}{c | c  | c | c}
e_0 & e_1 & \cdots & e_{m-1} \end{array} \right) $.
Vectors $ e_f $ and $ e_l $ denote the standard basis vectors with a 1 in the \underline{$f$}irst and \underline{$l$}ast position, respectively. The size of the vectors is determined by context.  
The zero matrix ``of appropriate size'' is denoted by $ 0 $.
\textcolor{review}{The number of rows of matrix $X$ is denoted as $m(X)$. Several symbols are used to denote the state of matrix partitions: $X^+$ for the future state at the end of an update step, $\widehat{X}$ for the original state, and $\widetilde{X}$ for partitions which have already been computed at a particular step. See \cite{vandegeijn2023derivingalgorithmstriangulartridiagonalization} for a comprehensive list.}

\subsection{Skew-symmetric (antisymmetric) matrices} \label{sec:skew}

\begin{definition}
    Matrix $ X \in \mathbb{R}^{m \times m} $ is said to be 
    \emph{skew-symmetric} if
    $ X = -X^T $.
\end{definition}

The diagonal elements of a skew-symmetric matrix equal zero and  $ \chi_{i,j} = - \chi_{j,i} $.

\begin{theorem}
Partition $ X  \in \mathbb{R}^{m \times m} $    
as
    $ \renewcommand{\arraystretch}{1.1}
    X = 
    \left( \begin{array}{c I c}
    X_{TL} & X_{TR} \\ \whline
    X_{BL} & X_{BR}
    \end{array} \right)$,
where $ X_{TL} $ is square.  Then $ X $ is skew-symmetric iff 
$
\left( \begin{array}{c I c}
    X_{TL} & X_{TR} \\ \whline
    X_{BL} & X_{BR}
    \end{array} \right)
    =
    \left( \begin{array}{c I c}
    -X_{TL}^T & -X_{BL}^T \\ \whline
    -X_{TR}^T & -X_{BR}^T
    \end{array} \right)$.
\end{theorem}

\NoShow{
\begin{theorem}
\label{thm:yTXy}
Let $ X\in \mathbb{R}^{m \times m} $ be a skew-symmetric matrix and $ y $ a vector of appropriate size.  Then $ y^T X y = 0 $.
\end{theorem}

\begin{proof}
    $ y^T X y
    = ( y^T X y )^T =
    y^T X^T y
    = y^T (-X) y = - y^T X y $.  Hence $ y^T X y = 0 $.
\end{proof}

\begin{corollary}
Let $ X $ be an $ m \times m $ skew-symmetric matrix and $ \chi_{i,j} $ its entries.  Then $ \chi_{i,i} = 0 $ for all $ 0 \leq i < m $.
\end{corollary}

\begin{proof}
    $ \chi_{i,i} = e_i^T X e_i = 0 $ by Theorem~\ref{thm:yTXy}.
\end{proof}
}

\begin{theorem}
Let  $ B \in \mathbb{R}^{m \times n}$.  If $ X\in \mathbb{R}^{n \times n} $ is skew-symmetric, then $ B X B^T $ is \textcolor{review2}{also}. 
\end{theorem}


The following theorem   connects a simpler blocked right-looking algorithm to blocked versions of  Wimmer's (two-step) unblocked  algorithm:

\begin{theorem}
\label{thm:TandS}
    Let matrix $ X $ be  skew-symmetric and assume there \textcolor{review}{exists} a matrix $ B $ and tridiagonal skew-symmetric matrix $ T $ such that $ X = B T B^T $.
    Then
    \begin{enumerate}
        \item 
        $ T = S - S^T$ where 
        \end{enumerate}
        \begin{equation}
        \renewcommand{\arraystretch}{1.0}
        \setlength{\arraycolsep}{3pt}
        \label{eqn:TvsS} T = \mbox{\footnotesize $\left( \begin{array}{c c c c c}
        0 & - \tau_{10} & 0 & 0 & \cdots \\ 
        \tau_{10}  & 0 & - \tau_{21} & 0 &  \cdots \\
        0 &  \tau_{21} &  0 & - \tau_{32} & \cdots  \\
        0 &  0 & \textcolor{review}{\tau_{32}} &  0 & \ddots \\
        \vdots & \vdots &  \vdots & \ddots & \ddots 
        \end{array}
        \right)$}
        \text{ and }
        S = \mbox{\footnotesize $\left( \begin{array}{c c c c c}
        0 & - \tau_{10} & 0 & 0 & \cdots \\ 
        0
        & 0 & 
        0
        & 0 &  \cdots \\
        0 &  \tau_{21} &  0 & - \tau_{32} & \cdots  \\
        0 &  0 & 
        0
        &  0 & \ddots \\
        \vdots & \vdots &  \vdots & \ddots & \ddots 
        \end{array}
        \right)$}.
        \end{equation}
    \begin{enumerate}
    \setcounter{enumi}{1}
        \item 
        $ X = B T B^T = B ( S - S^T ) B^T = 
        (B S)
        B^T - B
        (B S)^T
        =  W B^T - B W^T $,
        where $ W = B S $ has every other column equal to zero, starting with the first column.
    \end{enumerate}
\end{theorem}

\subsection{Gauss transforms}

We explain Gauss transforms as they are, for example,  used in the LU factorization of a matrix.

\NoShow{
The computation of the LU factorization
can be organized as the   application of a sequence of {\em Gauss transforms}:
If one
partitions 
\[
A = 
\left( \begin{array}{c | c} 
\alpha_{11} & a_{12}^T \\ \hline
a_{21} & A_{22}
\end{array} \right),
L =
\left( \begin{array}{c | c} 
1 & 0 \\ \hline
l_{21} & L_{22}
\end{array} \right),
\mbox{~and~}
U =
\left( \begin{array}{c | c} 
\upsilon_{11} & u_{12}^T \\ \hline
0 & U_{22}
\end{array} \right).
\]
then $ A = L U $ implies that
\[
\left( \begin{array}{c | c} 
\alpha_{11} & a_{12}^T \\ \hline
a_{21} & A_{22}
\end{array} \right)
=
\left( \begin{array}{c | c} 
1 & 0 \\ \hline
l_{21} & L_{22}
\end{array} \right)
\left( \begin{array}{c | c} 
\upsilon_{11} & u_{12}^T \\ \hline
0 & U_{22}
\end{array} \right)
=
\left( \begin{array}{c | c} 
\upsilon_{11} & u_{12}^T \\ \hline
\upsilon_{11} l_{21} & l_{21} u_{12}^T + L_{22} U_{22}
\end{array} \right).
\]
If we choose $ l_{21} = a_{21} / \alpha_{11} $, then
one updates
\[
\left( \begin{array}{c | c} 
\alpha_{11} & a_{12}^T \\ \hline
a_{21} & A_{22}
\end{array} \right)
:=
\left( \begin{array}{c | c} 
1 & 0 \\ \hline
-l_{21} & I
\end{array} \right)
\left( \begin{array}{c | c} 
\alpha_{11} & a_{12}^T \\ \hline
a_{21} & A_{22}
\end{array} \right)
=
\left( \begin{array}{c | c} 
\alpha_{11} & a_{12}^T \\ \hline
0 & A_{22} - l_{21} a_{12}^T
\end{array} \right).
\]
  Continuing this process with the updated $ A_{22} $ will ultimately overwrite $ A $ with $ U $ (provided $ A $ has nonsingular leading principle submatrices).
}

\begin{definition}
    A matrix 
    $
    L_{i} = 
    \left( \begin{array}{c I c | c}
    I_{i \times i} & 0 & 0 \\ \whline
    0 & 1 & 0 \\ \hline
    0 & l_{21}^{(i)} & I 
    \end{array} \right)
    $
    is called a \emph{Gauss transform}.
\end{definition} 
The inverse of a Gauss transform is also a Gauss transform:
\begin{lemma}
    $
    \left( \begin{array}{c I c | c}
    I_{i \times i} & 0 & 0 \\ \whline
    0 & 1 & 0 \\ \hline
    0 & l_{21}^{(i)} & I 
    \end{array} \right)^{-1}
    =
    \left( \begin{array}{c I c | c}
    I_{i \times i} & 0 & 0 \\ \whline
    0 & 1 & 0 \\ \hline
    0 & - l_{21}^{(i)} & I 
    \end{array} \right)$. 
\end{lemma}
\NoShow{With this, the described process for computing the LU factorization can be summarized as
\[
L_{n-1}^{-1} \cdots L_1^{-1} L_0^{-1} A = U
\mbox{~or, equivalently,~}
A = 
L_0 L_1 \cdots L_{n-1}
U = L U,
\]
where each $ L_i 
$ is a Gauss transform with appropriately chosen $ l_{21}^{(i)}  $.  }
The product of Gauss transforms, $ L_0 L_1 \cdots L_{n-1} $, is a unit lower-triangular matrix $ L $ that 
consists of the identity 
with $ l_{21}^{(i)} $ of $ L_i $  inserted in the column indexed with $ i $:
\begin{theorem}
If the matrices in the following expression are conformally partitioned, then
$
\begin{array}[t]{c}
\underbrace{
\left( \begin{array}{c I c | c}
L_{00} & 0 & 0 \\ \whline
l_{10}^T & 1 & 0 \\ \hline
L_{20} & 0 & I 
\end{array} \right)
}
\\
L_0 \cdots L_{i-1}
\end{array}
\begin{array}[t]{c}
\underbrace{
\left( \begin{array}{c I c | c}
I & 0 & 0 \\ \whline
0 & 1 & 0 \\ \hline
0 & l_{21}^{(i)} & I 
\end{array} \right)
} \\
L_{i}
\end{array}
=
\begin{array}[t]{c}
\underbrace{
\left( \begin{array}{c I c | c}
L_{00} & 0 & 0 \\ \whline
l_{10}^T & 1 & 0 \\ \hline
L_{20} & l_{21}^{(i)} & I 
\end{array} \right)
}
\\
L_0 \cdots L_{i-1} L_{i}
\end{array}$.
\end{theorem}
This means that
$
L_0 L_1 \cdots L_{n-1}
=
\left( \begin{array}{c | c | c | c}
1 & 0 & 0 & 0 \\ \hline
\multirow{3}{*}{
$l_{21}^{(0)}$} & 1 & 0 & \cdots \\ \cline{2-4}
{} &  \multirow{2}{*}{
$l_{21}^{(1)}$} & 1 & \cdots  \\ \cline{3-4}
{} & {} & l_{21}^{(2)}& \ddots 
\end{array}
\right)$.
Thus, a matrix of the form
$ \renewcommand{\arraystretch}{1.1}
\left( \begin{array}{ c | c}
L_{TL} & 0 \\ \hline
L_{BL} &  I 
\end{array} \right)
$,
where $ L_{TL} $ is unit lower triangular, is a \emph{block Gauss transform} that
represents an accummulation of Gauss transforms.
\NoShow{
This, and the following corollary,  will  play a critical role in the development of so-called blocked algorithms that cast most computation in terms of matrix-matrix multiplication.
\begin{corollary}
\label{cor:inv}
$
    \left( \begin{array}{ c | c}
    L_{TL} & 0 \\ \hline
    L_{BL}  &  I
    \end{array} \right)^{-1}
    \left( \begin{array}{ c | c}
    L_{TL} & 0 \\ \hline
    L_{BL}  &  L_{BR} 
    \end{array} \right)
    =
    \left( \begin{array}{ c | c}
    I & 0 \\ \hline
    0  &  L_{BR} 
    \end{array} \right)
$.
\end{corollary}

\begin{proof}
    The result follows immediately from the observation that
    \[
     \left( \begin{array}{ c | c}
    L_{TL} & 0 \\ \hline
    L_{BL}  &  I
    \end{array} \right)^{-1}
    =
     \left( \begin{array}{ c | c}
    L_{TL}^{-1} & 0 \\ \hline
    - L_{BL} L_{TL}^{-1} &  I
    \end{array} \right).
    \]
\end{proof}
}

\NoShow{ 
\begin{corollary}
    If 
    $ L_0 L_1 \cdots L_{b-1} =
    \left( \begin{array}{ c | c}
    L_{11} & 0 \\ \hline
    L_{21} &  I 
    \end{array} \right)
    $ then
    \[
    L_{b-1}^{-1} \cdots L_1^{-1} L_0^{-1} =
    \left( \begin{array}{ c | c}
    L_{11} & 0 \\ \hline
    L_{21} &  I 
    \end{array} \right)^{-1}
    =
    \left( \begin{array}{ c | c}
    L_{11}^{-1} & 0 \\ \hline
    - L_{21} L_{11}^{-1} &  I 
    \end{array} \right)
    =
    \left( \begin{array}{ c | c}
    I & 0 \\ \hline
    - L_{21}  &  I 
    \end{array} \right)
    \left( \begin{array}{ c | c}
    L_{11}^{-1} & 0 \\ \hline
    0 &  I 
    \end{array} \right)
    .
    \]
\end{corollary}
} 

\NoShow{
\begin{theorem}
\label{thm:right-update}
  Let $ C = \left( 
    \begin{array}{c | c}
    C_{TL} & - C_{BL}^T \\ \hline
    C_{BL} & C_{BR} 
    \end{array}
    \right) $ be skew-symmetric and $ B = \left( 
    \begin{array}{c | c}
    B_{TL} & 0 \\ \hline
    B_{BL} & I 
    \end{array}
    \right) $ be nonsingular (e.g., a unit lower-triangular matrix).  Consider the equality  $ C^{\rm +} = B^{-1}  C B^{-T} $ in terms of quadrants.  Then
    \[
    \left( 
    \begin{array}{c | c}
    C_{TL}^{\rm +} & - C_{BL}^{T \! \rm +} \\ \hline
    C_{BL}^{\rm +} & C_{BR}^{\rm +} 
    \end{array}
    \right)
    =
    \left( 
    \begin{array}{c | c}
    B_{TL} & 0 \\ \hline
    B_{BL} & I 
    \end{array}
    \right)^{-1}
    \left( 
    \begin{array}{c | c}
    C_{TL} & - C_{BL}^T \\ \hline
    C_{BL} & C_{BR} 
    \end{array}
    \right)
    \left( 
    \begin{array}{c | c}
    B_{TL} & 0 \\ \hline
    B_{BL} & I 
    \end{array}
    \right)^{-T}  
    \]
    implies that
    \begin{itemize}
        \item 
        $ C_{TL}^{\rm +} = B_{TL}^{-1} C_{TL} B_{TL}^{-T} $.
        \item 
        $ 
        C_{BL}^{\rm +} = - B_{BL} C_{TL}^{\rm +} 
        + C_{BL} B_{TL}^{-T}$.
        \item 
        $  C_{BR}^{\rm +}
    =
    C_{BR} - 
    B_{BL} C_{TL}^{\rm +} B_{TR}^T + ( B_{BL}  C_{BL}^{{\rm +} \!T}
    -
    C_{BL}^{\rm +} B_{TR}^T )  
        $.

    \end{itemize}
\end{theorem}

\begin{proof}
\[
\begin{array}{l}
    \left( 
    \begin{array}{c | c}
    C_{TL} & - C_{BL}^T \\ \hline
    C_{BL} & C_{BR} 
    \end{array}
    \right)
    =
\left( 
    \begin{array}{c | c}
    B_{TL} & 0 \\ \hline
    B_{BL} & I 
    \end{array}
    \right)
\left( 
    \begin{array}{c | c}
    C_{TL}^{\rm +} & - C_{BL}^{{\rm +} \!T } \\ \hline
    C_{BL}^{\rm +} & C_{BR}^{\rm +} 
    \end{array}
    \right)
\left( 
    \begin{array}{c | c}
    B_{TL}^T & B_{BL}^T \\ \hline
    0 & I 
    \end{array}
    \right) \\
    \NoShow{
~~ =
\left( 
    \begin{array}{c | c}
    B_{TL} & 0 \\ \hline
    B_{BL} & I 
    \end{array}
    \right)
 \left( 
    \begin{array}{c | c}
    C_{TL}^{\rm +} B_{TL}^T & 
    C_{TL}^{\rm +} B_{TR}^T - C_{BL}^{{\rm +} \!T } \\ \hline
    C_{BL}^{\rm +} B_{TL}^T  & C_{BL}^{\rm +} B_{TR}^T + C_{BR}^{\rm +} 
    \end{array}
    \right)   \\
    } 
~~ =
 \left( 
    \begin{array}{c | c}
    B_{TL} C_{TL}^{\rm +} B_{TL}^T & 
      \star \\ \hline
    B_{BL} C_{TL}^{\rm +} B_{TL}^T +
    C_{BL}^{\rm +} B_{TL}^T  & 
    B_{BL} C_{TL}^{\rm +} B_{TR}^T - B_{BL}  C_{BL}^{{\rm +} \!T}
    +
    C_{BL}^{\rm +} B_{TR}^T + C_{BR}^{\rm +} 
    \end{array}
    \right)   
    \end{array}
\]
from  which the desired results follow via straight-forward manipulation.
\begin{itemize}
    \item 
    $ C_{TL}^{\rm +} = B_{TL}^{-1} C_{TL} B_{TL}^{-T} $.
    \item 
    $ C_{BL}^{\rm +} = ( C_{BL} - B_{BL} C_{TL}^{\rm +} B_{TL}^T )  B_{TL}^{-T}=
    C_{BL}B_{TL}^{-T} - B_{BL} C_{TL}^{\rm +}
    $
    \item 
    $ C_{BR}^{\rm +}
    =
    C_{BR} - 
    B_{BL} C_{TL}^{\rm +} B_{TR}^T + B_{BL}  C_{BL}^{{\rm +} \!T}
    -
    C_{BL}^{\rm +} B_{TR}^T   
    $
\end{itemize}
\end{proof}
} 

\begin{figure}[tb!]
    \input LTLt_piv_unb.tex
\centering
    
    \footnotesize
    \FlaAlgorithm    
    \caption{The unblocked right-looking (modified Parlett-Reid) and left-looking (modified Aasen) algorithms. Portions in red are only required when pivoting is desired. See Section~\ref{sec:pivoting} for definitions related to pivoting.}
    \label{fig:LTLt_unb}

\end{figure}

\subsection{Modifying the Parlett-Reid algorithm}
\label{sec:simple}

With these tools, we describe 
an algorithm for computing the $ L T L^T $ factorization for a skew-symmetric matrix, modified from one first proposed by Parlett and Reid for symmetric matrices~\cite{ParlettReid}.

Partition
$
X \rightarrow
\left( \begin{array}{c | c | c}
\chi_{11} & - \chi_{21} & - x_{31}^T \\ \hline
\chi_{21} &  \chi_{22} & - x_{32}^T \\ \hline
x_{31} & x_{32} & X_{33}
\end{array} \right)
$.
The \textcolor{review}{objective is then} to find a Gauss transform to introduce zeroes in $ x_{31} $:
\begin{equation}
\label{eqn:1}
\begin{array}{l}
\left( \begin{array}{c | c | c}
\chi_{11} & - \chi_{21} & 0 \\ \hline
\chi_{21} &  \chi_{22}^+ & -  x_{32}^{{\rm +} \! T} \\ \hline
0 & x_{32}^{\rm +} &  X_{33}^{\rm +}
\end{array} \right)
:=  \\
~~~~~~~
\left( \begin{array}{c | c | c}
1 & 0 & 0 \\ \hline
0 & 1 & 0 \\ \hline
0 & -l_{32} & I
\end{array} \right)
\left( \begin{array}{c | c | c}
\chi_{11} & - \chi_{21} & - x_{31}^T \\ \hline
\chi_{21} &  \chi_{22} & - x_{32}^T \\ \hline
x_{31} & x_{32} & X_{33}
\end{array} \right)
\left( \begin{array}{c | c | c}
1 & 0 & 0 \\ \hline
0 & 1 & -l_{32}^T \\ \hline
0 & 0 & I
\end{array} \right)
.
\end{array}
\end{equation}
Here, the submatrices with a $+$ superscript equal the contents of  the indicated parts of the matrix after the update (application of Gauss transforms from left and right).
Equation~(\ref{eqn:1}) suggests
updating
\begin{itemize}
    \item 
    $ l_{32} := x_{31} / \chi_{21} $. 
    \item 
    $ x_{31}:= 0 $.  
    \item
    \setlength{\arraycolsep}{3pt}
    $ 
    \begin{array}[t]{@{}r@{~}c@{~}l}
    \left( \begin{array}{c | c}
    \chi_{22} & -x_{32}^T \\ \hline
    x_{32} & X_{33}
    \end{array} \right)
    &:=&
    \left( \begin{array}{c | c}
    1 & 0 \\ \hline
    - l_{32} & I 
    \end{array} \right)
    \left( \begin{array}{c | c}
    \chi_{22} & - x_{32}^T \\ \hline
    x_{32} & X_{33} 
    \end{array} \right)
     \left( \begin{array}{c | c}
    1 & - l_{32}^T \\ \hline
    0  & I 
    \end{array} \right) \\
   & =&
    \left( \begin{array}{c | c}
    \chi_{22} & - x_{32}^T \\ \hline
    x_{32} & X_{33} + ( l_{32} x_{32}^T
    - x_{32} l_{32}^T )
    \end{array} \right).
    \end{array}
    $
    \item
    Continue the factorization with the updated
    $
    \left( \begin{array}{c | c}
    \chi_{22} & - x_{32}^T \\ \hline
    x_{32} & X_{33}
    \end{array} \right)$.

\end{itemize}

    Note that the skew-symmetric structure of $X$ is \emph{implicit}, that is, upper triangular portions of skew-symmetric matrices or sub-matrices are neither stored nor referenced, such as $x_{32}^T$ or the upper triangle of $X_{33}$. In practice, $ X_{33} $ is updated by a skew-symmetric rank-2 update which respects this implicit structure. Importantly, this also means that the diagonal elements of $X$, such as $\chi_{22}$, are \emph{assumed} to be zero and not actually computed, like the diagonal elements of $L$ are  ones but not stored (except as noted later for implementation reasons). Due to this implicit structure, we do not reference any unstored elements in the presented algorithms.
    
The resulting algorithm, in FLAME notation, is given in Fig.~\ref{fig:LTLt_unb}.
The partitioning and repartitioning in that algorithm is consistent with the use of the thin lines and the choice of subscripting  earlier in this section.  
\NoShow{
Matlab code is given in Fig.~\ref{fig:LTLt_unb_right_matlab}.
}

\NoShow{
Equation (\ref{eqn:1}) can be rearranged as
\[
\left( \begin{array}{c | c | c}
0 & - \chi_{21} & - x_{31}^T \\ \hline
\chi_{21} &  0 & - x_{32}^T \\ \hline
x_{31} & x_{32} & X_{33}
\end{array} \right)
=
\begin{array}[t]{c}
\underbrace{
\left( \begin{array}{c | c | c}
1 & 0 & 0 \\ \hline
0 & 1 & 0 \\ \hline
0 & l_{31} & I
\end{array} \right)
} \\
L_0 
\end{array}
\left( \begin{array}{c | c | c}
0 & - \chi_{21} & 0 \\ \hline
\chi_{21} &  0 & - \widehat x_{32}^T \\ \hline
0 & \widehat x_{32} & \widehat X_{33}
\end{array} \right)
\begin{array}[t]{c}
\underbrace{
\left( \begin{array}{c | c | c}
1 & 0 & 0 \\ \hline
0 & 1 & l_{31}^T \\ \hline
0 & 0 & I
\end{array} \right)
}
\\
L_0^T
\end{array}.
\]
After $ n-1 $ iterations, we find that
\[
X
=
L_0 
\cdots
L_{n-2} 
T
L_{n-2}^T 
\cdots
L_0^T,
\]
where $ T $ is a tridiagonal skew-symmetric matrix.

The observation that
\begin{equation}
    \label{eqn1:L1}
\left( \begin{array}{c I c | c}
L_{00} & 0 & 0 \\ \whline
l_{10} & 1 & 0 \\ \hline
L_{20} & 0 & I 
\end{array} \right)
\left( \begin{array}{c I c | c}
I & 0 & 0 \\ \whline
0 & 1 & 0 \\ \hline
0 & l_{21} & I 
\end{array} \right)
=
\left( \begin{array}{c I c | c}
L_{00} & 0 & 0 \\ \whline
l_{10} & 1 & 0 \\ \hline
L_{20} & l_{21} & I 
\end{array} \right)
\end{equation}
tells us each encountered
$ \left( \begin{array}{c}
0 \\ \hline
1 \\ \hline
l_{31}= 
\end{array} \right) $ becomes 
the corresponding column of a unit lower-triangular matrix $ L $ so that 
\[
X =
L X L^T =
\left( \begin{array}{c | c}
1 & 0 \\ \hline
0 & \widetilde L
\end{array} \right)
T
\left( \begin{array}{c | c}
1 & 0 \\ \hline
0 & \widetilde L^T
\end{array} \right),
\]
where $ L $ and $ \widetilde L $ are unit lower triangular matrices.
The strictly lower triangular part of $ L $ overwrites $ X $ below the first subdiagonal.  In other words, in Matlab notation,
\begin{center}
    {\tt Ltilde = tril( X( 2:n, 1:n-1 ), -1 ) + eye( n-1, n-1 )}.
\end{center}
}

\section{A family of algorithms}\label{sec:flame}

We now outline how multiple  algorithms can be systematically derived from specifications.

\subsection{The FLAME workflow}


   We  briefly review how the FLAME methodology systematically discovers  families of algorithms.
   This, followed by the translation 
   into code using a FLAME API, is what we refer to as  the {\em FLAME  workflow}.
   
 The  FLAME notation (hiding explicit indexing) 
     enables the rapid discovery of DLA algorithms via the application of formal derivation techniques~\cite{FLAME_WoCo,FLAME_TR,Recipe,TSoPMC}.
   \textcolor{review}{In the FLAME workflow,} one starts with
   the  operation
   \textcolor{review}{(Sec.~\ref{sec:spec})}
   from which the 
   Partitioned Matrix Expression (PME)
   (a recursive definition of the operation,
   \textcolor{review}{Sec.~\ref{sec:pme})} is obtained.  From this,
   loop invariants (logical conditions that \textcolor{review}{capture} the state of variables before and after each iteration,
   \textcolor{review}{Sec.~\ref{sec:invariants})}
   can be deduced. \textcolor{review}{A \emph{worksheet outline} is then filled,}  deriving (hand in hand  with their proofs of correctness) 
   algorithmic variants
   \textcolor{review}{(Sec.~\ref{sec:unblocked}, \ref{sec:blk}).}
    Families of algorithms for a broad range of DLA operations (within and beyond LAPACK) have been systematically derived~\cite{FLAME,TSoPMC,Recipe,Bientinesi:2008:FAR:1377603.1377606,10.1145/3544585.3544597}. 
    By adopting APIs that mirror the FLAME notation, correct algorithms are translated 
    to correct code
    e.g. the FLAMEC API used by the  libflame DLA library~\cite{libflame_github,libflamebook,CiSE09}.

\subsection{
Specification} \label{sec:spec}

Given a skew-symmetric matrix $ X $,
the goal is to compute a unit lower triangular matrix $ L $ and tridiagonal matrix $ T $ such that $ X = L T L^T $, overwriting $ X $ with $ T $, provided such a factorization exists.
We specify this with the \emph{precondition} 
$
X = \widehat X \wedge ( \exists L, T ~ \vert ~ \widehat X = L T L^T ) 
$
and \emph{postcondition} 
$ X = T \wedge \widehat X = L T L^T $, where $\widehat X $ equals the original contents of $ X $ and the special structures of the various matrices are implicit.
Since in practice the strictly lower triangular part of $ L $ typically overwrites the entries below the first subdiagonal of $ T $ (although this is not asserted by our postcondition), the first column of $ L $ equals $ e_0 $.  However, as was pointed out in~\cite{Miroslav2011}, this is only one choice for the first column of $ L $.
Indeed, if
\[
\setlength{\arraycolsep}{2pt}
\begin{array}[t]{c}
\underbrace{
\left( \begin{array}{ c | c}
0 & \star \\ \hline
\widehat x_{21} & \widehat X_{22}
\end{array} \right)
} \\
\widehat X
\end{array}
=
\begin{array}[t]{c}
\underbrace{
\left( \begin{array}{c | c}
1 & 0 \\ \hline
l_{21} & L_{22}
\end{array}
\right)
} \\
L
\end{array}
\begin{array}[t]{c}
\underbrace{
\left( \begin{array}{c | c}
0 & \star \\ \hline
\tau_{21} e_f & T_{22}
\end{array}
\right)
} \\
T
\end{array}
\begin{array}[t]{c}
\underbrace{
\left( \begin{array}{c | c}
1 & 0 \\ \hline
l_{21} & L_{22}
\end{array}
\right)^T
} \\
L^T
\end{array}
\]
for some choice of $ l_{21} $, 
then 
\[
\setlength{\arraycolsep}{2pt}
\left( \begin{array}{c | c}
1 & 0 \\ \hline
- l_{21} & I
\end{array}
\right)
\left( \begin{array}{ c | c}
0 & \star \\ \hline
\widehat x_{21} & \widehat X_{22}
\end{array} \right)
\left( \begin{array}{c | c}
1 & 0 \\ \hline
- l_{21} & I
\end{array}
\right)^T =
\left( \begin{array}{c | c}
1 & 0 \\ \hline
0 & L_{22}
\end{array}
\right)
\left( \begin{array}{c | c}
0 & \star \\ \hline
\tau_{21} e_f & T_{22}
\end{array}
\right)
\left( \begin{array}{c | c}
1 & 0 \\ \hline
0 & L_{22}
\end{array}
\right)^T,
\]
which means the original matrix $ X $ can always be updated by applying the first Gauss transform, defined by $ l_{21} $, from the left and right or, equivalently, 
$
X_{22} := 
X_{22} + ( l_{21} x_{21}^T - x_{21} l_{21}^T )
$,
before executing the algorithm given in Section~\ref{sec:simple}.

\subsection{Deriving the Partitioned Matrix Expession} \label{sec:pme}

We derive the PME from the specification of the operation by substituting partitioned matrices into the postcondition.
For most DLA factorization algorithms, 
matrices were partitioned into quadrants.  When the methodology was applied to 
Krylov subspace methods~\cite{Eijkhout20101805}, where upper Hessenberg and tridiagonal matrices are encountered, $ 3 \times 3 $ partitionings were necessary.  Not surprisingly, 
given the algorithm presented in Fig.~\ref{fig:LTLt_unb}, this is also found to be the case when deriving algorithms for the $ L T L^T $ factorization.

Thus, for the PME we find
{\footnotesize
\begin{eqnarray}
\nonumber
\lefteqn{
\left( \begin{array}{c I c | c}
X_{TL} &  \star
& \star \\ \whline
x_{ML}^T & 0 & \star \\ \hline
X_{BL} & x_{BM} & X_{BR}
\end{array} \right)
=
\left( \begin{array}{c I c | c}
T_{TL} & \star & \star \\ \whline
\tau_{ML} e_l^T & 0 &  \star \\ \hline
0  & \tau_{BM} e_f & T_{BR}
\end{array} \right) \wedge  
\left( \begin{array}{c I c | c}
\widehat X_{TL} & 
\star & 
\star \\ \whline
\widehat x_{ML}^T & 0 & 
\star \\ \hline
\widehat X_{BL} & \widehat x_{BM} & \widehat X_{BR}
\end{array} \right) =}
\\ 
\nonumber
& 
 &
\left( \begin{array}{c I c | c}
L_{TL} & 0 & 0 \\ \whline
l_{ML}^T & 1 & 0 \\ \hline
L_{BL} & l_{BM} &  L_{BR}
\end{array}\right)
\left( \begin{array}{c I c | c}
T_{TL} & \star  & \star \\ \whline
\tau_{ML} e_l^T  & 0 & \star \\ \hline
0 & \tau_{BM} e_f & T_{BR}
\end{array} \right)
\left( \begin{array}{c I c | c}
L_{TL}^T & l_{ML} & L_{BL}^T \\ \whline
0 & 1 & l_{BM}^T \\ \hline
0 & 0 &  L_{BR}^T
\end{array}\right).
\end{eqnarray}
}%
The $ \star $s capture that those expressions are not stored due to implicit skew-symmetry.
\comment{
The right hand side of the second condition can be rewritten as
{\footnotesize
\begin{eqnarray*}
\NoShow{\lefteqn{
\left( \begin{array}{c I c | c}
X_{TL} & \star & \star \\ \whline
x_{ML}^T & \chi_{MM} & \star \\ \hline
X_{BL} & x_{BM} & X_{BR}
\end{array} \right)
=
\left( \begin{array}{c I c | c}
T_{TL} & \star & \star \\ \whline
\tau_{ML} e_l^T & 0 &  \star \\ \hline
0  & \tau_{BM} e_f & T_{BR}
\end{array} \right) \wedge 
\left( \begin{array}{c I c | c}
\widehat X_{TL} & \star & \star \\ \whline
\widehat x_{ML}^T & 0 & \star \\ \hline
\widehat X_{BL} & \widehat x_{BM} & \widehat X_{BR}
\end{array} \right)} \\
} 
\left( \begin{array}{c I c | c}
L_{TL} & 0 & 0 \\ \whline
l_{ML}^T & 1 & 0 \\ \hline
L_{BL} & l_{BM} &  I
\end{array}\right)
\left( \begin{array}{c I c | c}
T_{TL} & - \tau_{ML} e_l  & 0 \\ \whline
\tau_{ML} e_l^T  & 0 & -\tau_{BM} (L_{BR} e_f)^T \\ \hline
0 & \tau_{BM} L_{BR} e_f & L_{BR} T_{BR} L_{BR}^T
\end{array} \right)
\left( \begin{array}{c I c | c}
L_{TL}^T & l_{ML} & L_{BL}^T \\ \whline
0 & 1 & l_{BM}^T \\ \hline
0 & 0 &  I
\end{array}\right).
\end{eqnarray*}
}%
Here
\[
\left( \begin{array}{c | c}
0 & - \tau_{BM} (L_{BR}  e_f)^T \\ \hline
\tau_{BM} L_{BR}  e_f & L_{BR} T_{BR} L_{BR}^T 
\end{array} \right)
=
\begin{array}[t]{c}
\underbrace{
\left( \begin{array}{c | c}
1 & 0 \\ \hline
0 & L_{BR} 
\end{array} \right)
} \\
L_{k} \cdots L_{m-2}
\end{array}
\left( \begin{array}{c | c}
0 & - \tau_{BM} e_f^T \\ \hline
\tau_{BM} e_f & T_{BR} 
\end{array} \right)
\begin{array}[t]{c}
\underbrace{
\left( \begin{array}{c | c}
1 & 0 \\ \hline
0 & L_{BR} 
\end{array} \right)^T,
} \\
L_{m-2}^T \cdots L_{k}^T
\end{array}
\]
which captures that it represents the result at a particular intermediate stage of the calculation
expressed as  the final result but with the yet-to-be-computed transformations not yet applied.%
\footnote{The exact number of Gauss transforms applied at a given step is tricky to account for due to the offset in $L$, leading to infamous ``off by one'' errors.  This becomes inconsequential since we avoid indices in our subsequent reasoning.}
This insight will play an important role in our derivation and deviates from how the FLAME methodology has been traditionally deployed. 
}

\subsection{Loop invariants} \label{sec:invariants}

A loop invariant is a predicate that captures the state of the variables before and after each iteration of the loop.
The strength of the FLAME methodology is that this condition is derived {\em a priori} from the PME so that it can guide the derivation of the loop.
From the PME, taking into account that we eventually wish to add pivoting, we find  the following 
loop invariants%
\footnote{There may be other loop invariants.}%
:

\subsubsection{Invariant~1 (right-looking)}
    { 
\footnotesize
\setlength{\arraycolsep}{2pt}
\begin{subequations}
\begin{align}\label{eqn:inv_unb_right_1}
&
\left( \begin{array}{c I c | c}
X_{TL} & \star & \star \\ \whline
x_{ML}^T & 0 & \star \\ \hline
X_{BL} & x_{BM} & X_{BR}
\end{array} \right)
=
\left( \begin{array}{c I c | c}
T_{TL} & \star & \star \\ \whline
 \tau_{ML} e_l^T & 0 &  \star \\ \hline
0  &  \tau_{BM} 
L_{BR} e_f
& L_{BR} T_{BR} L_{BR}^T
\end{array} \right)  \wedge \\
\label{eqn:inv_unb_right_2}
&\quad\quad
\left( \begin{array}{c I c | c}
\widehat X_{TL} & 
\star
& 
\star
\\ \whline
\widehat x_{ML}^T & 0 & 
\star \\ \hline
\widehat X_{BL} & \widehat x_{BM} & \widehat X_{BR}
\end{array} \right) = \\
\nonumber
&\quad\quad\quad\quad
\left( \begin{array}{c I c | c}
\widetilde{L}_{TL} & 0 & 0 \\ \whline
\widetilde{l}_{ML}^T & 1 & 0 \\ \hline
\widetilde{L}_{BL} & \widetilde{l}_{BM} &  I
\end{array}\right)
\left( \begin{array}{c I c | c}
T_{TL} & \star  & \star \\ \whline
\tau_{ML} e_l^T  & 0 & \star \\ \hline
0 & \tau_{BM} L_{BR} e_f & L_{BR} T_{BR} L_{BR}^T
\end{array} \right)
\left( \begin{array}{c I c | c}
\widetilde{L}_{TL}^T & \widetilde{l}_{ML} & \widetilde{L}_{BL}^T \\ \whline
0 & 1 & \widetilde{l}_{BM}^T \\ \hline
0 & 0 &  I
\end{array}\right).
\end{align}
\end{subequations}
    }%
    \NoShow{
    \item 
    Invariant for  a variation on the right-looking algorithm, which we will call lazy right-looking, that in a current iteration computes the next Gauss transform but does not yet apply it: 
{
\setlength{\arraycolsep}{2pt}
\[
\begin{array}{l}
\left( \begin{array}{c I c | c}
X_{TL} & \star & \star \\ \whline
x_{ML}^T & \chi_{MM} & \star \\ \hline
X_{BL} & x_{BM} & X_{BR}
\end{array} \right)
=
\left( \begin{array}{c I c | c}
T_{TL} & ~~~~~~~\star ~~~~~~~& \star \\ \whline
 \tau_{ML} e_l^T & 
 \multicolumn{2}{c}{
 \multirow{2}{*}{
 $
 \left( \begin{array}{c | c}
 1 &  0 \\ \hline
l_{BM} & L_{BR} 
\end{array}
\right)
 \left( \begin{array}{c | c}
 0 &  \star \\ \hline
\tau_{BM} 
e_f
& T_{BR} 
\end{array}
\right)
\left( \begin{array}{c | c}
 1 &  0 \\ \hline
l_{BM} & L_{BR} 
\end{array}
\right)^T
$
 }
 }
  \\ \cline{1-1}
0  &  \multicolumn{2}{c}{}
\end{array} \right)
\wedge \\
\\[-0.15in]
~~ 
\left( \begin{array}{c I c | c}
\widehat X_{TL} & \star & \star \\ \whline
\widehat x_{ML}^T & 0 & \star \\ \hline
\widehat X_{BL} & \widehat x_{BM} & \widehat X_{BR}
\end{array} \right)  =
\left( \begin{array}{c I c | c}
\widetilde{L}_{TL} & 0 & 0 \\ \whline
\widetilde{l}_{ML}^T & 1 & 0 \\ \hline
\widetilde{L}_{BL} & \widetilde{l}_{BM} &  L_{BR}
\end{array}\right)
\left( \begin{array}{c I c | c}
T_{TL} & - \tau_{ML} e_l  & 0 \\ \whline
\tau_{ML} e_l^T  & 0 & -\tau_{BM} e_f^T \\ \hline
0 & \tau_{BM} e_f & T_{BR}
\end{array} \right)
\left( \begin{array}{c I c | c}
\widetilde{L}_{TL}^T & \widetilde{l}_{ML} & \widetilde{L}_{BL}^T \\ \whline
0 & 1 & \widetilde{l}_{BM}^T \\ \hline
0 & 0 &  L_{BR}^T
\end{array}\right).
\end{array}
\]
}%
}%
Only the parts of $ L $ \textcolor{review}{denoted with a tilde (e.g. $\widetilde{L}_{BL}$)} have been computed.
One additional column of $L$ is known at a given step beyond the position of the ``thick line''. This is because the Gauss transform vector is computed from the previous column of $ X $, as seen in Section~\ref{sec:simple}.
Finally, in~(\ref{eqn:inv_unb_right_1})
{
\footnotesize
\[
\left( \begin{array}{c | c}
0 & \star \\ \hline
\tau_{BM} L_{BR}  e_f & L_{BR} T_{BR} L_{BR}^T 
\end{array} \right)
= 
\begin{array}[t]{c}
\underbrace{
\left( \begin{array}{c | c}
1 & 0 \\ \hline
0 & L_{BR} 
\end{array} \right)
} \\
L_{k} \cdots L_{m-2}
\end{array}
\left( \begin{array}{c | c}
0 & \star \\ \hline
\tau_{BM} e_f & T_{BR} 
\end{array} \right)
\begin{array}[t]{c}
\underbrace{
\left( \begin{array}{c | c}
1 & 0 \\ \hline
0 & L_{BR} 
\end{array} \right)^T
} \\
L_{m-2}^T \cdots L_{k}^T
\end{array}
\]
}%
represents the result at a particular intermediate stage of the calculation \textcolor{review}{where the final block of Gauss transforms (represented by $L_{BR}$) have not yet been applied. $L_{BR}$ appears in the expression as an inverse transformation which ``undoes'' the transformations implicit in the final tridiagonal block $T_{BR}$.}
\footnote{The exact number of Gauss transforms applied at a given step is tricky to account for due to the offset in $L$. 
This becomes inconsequential since we avoid indices in our subsequent reasoning.}
This formulation is a departure from how invariants have been traditionally presented in the FLAME methodology and plays an important role in the detail of the derivations in~\cite{vandegeijn2023derivingalgorithmstriangulartridiagonalization}. 

\subsubsection
{Invariant~2a (fused right-looking)}
    {
    \footnotesize
\setlength{\arraycolsep}{2pt}

\[
\left( \begin{array}{c I c | c}
X_{TL} & \star & \star \\ \whline
x_{ML}^T & 0 & \star \\ \hline
X_{BL} & x_{BM} & X_{BR}
\end{array} \right)
=
\left( \begin{array}{c I c | c}
T_{TL} & ~~~~~~~~~~~~~~\star ~~~~~~~~~~~~~~& \star \\ \whline
 \tau_{ML} e_l^T & 
 \multicolumn{2}{c}{
 \multirow{2}{*}{
 $
 \left( \begin{array}{c | c}
 1 &  0 \\ \hline
\widetilde{l}_{BM} & L_{BR} 
\end{array}
\right)
 \left( \begin{array}{c | c}
 0 &  \star \\ \hline
\tau_{BM} 
e_f
& T_{BR} 
\end{array}
\right)
\left( \begin{array}{c | c}
 1 &  \widetilde{l}_{BM}^T \\ \hline
0 & L_{BR}^T
\end{array}
\right)
$
 }
 }
  \\ \cline{1-1}
0  &  \multicolumn{2}{c}{}
\end{array} \right) 
\wedge \mbox{
(\ref{eqn:inv_unb_right_2})}.
\]
    }%
    This invariant differs  from the  first  by capturing that the last Gauss transform (stored in $ \widetilde{l}_{BM} $) has been computed but not yet applied to the rest of matrix $ X $.

    \subsubsection{Invariant~2b (fused right-looking)}
    { 
\footnotesize
\setlength{\arraycolsep}{2pt}
  \begin{eqnarray*}
  \nonumber
  \lefteqn{
\left( \begin{array}{c I c | c}
X_{TL} & \star & \star \\ \whline
x_{ML}^T & 0 & \star \\ \hline
X_{BL} & x_{BM} & X_{BR}
\end{array} \right)
=
\left( \begin{array}{c I c | c}
T_{TL} & \star & \star \\ \whline
 \tau_{ML} e_l^T & 0 &  \star \\ \hline
0  &  \tau_{BM} 
 e_f
& L_{BR} T_{BR} L_{BR}^T
\end{array} \right)  \wedge
\left( \begin{array}{c I c | c}
\widehat X_{TL} & 
\star
& 
\star
\\ \whline
\widehat x_{ML}^T & 0 & 
\star \\ \hline
\widehat X_{BL} & \widehat x_{BM} & \widehat X_{BR}
\end{array} \right) = }
\\
&& ~~~~ 
\left( \begin{array}{c I c | c}
\widetilde{L}_{TL} & 0 & 0 \\ \whline
\widetilde{l}_{ML}^T & 1 & 0 \\ \hline
\widetilde{L}_{BL} & \widetilde{l}_{BM} &  I
\end{array}\right)
\left( \begin{array}{c I c | c}
T_{TL} & \star  & \star \\ \whline
\tau_{ML} e_l^T  & 0 & \star \\ \hline
0 & \tau_{BM} \widetilde{L}_{BR} e_f & L_{BR} T_{BR} L_{BR}^T
\end{array} \right)
\left( \begin{array}{c I c | c}
\widetilde{L}_{TL}^T & \widetilde{l}_{ML} & \widetilde{L}_{BL}^T \\ \whline
0 & 1 & \widetilde{l}_{BM}^T \\ \hline
0 & 0 &  I
\end{array}\right).
\end{eqnarray*}
    }
This invariant differs from Invariant~1 in that it computes one more column of $ T $ and $ L $ (namely $ \widetilde{L}_{BR} e_f $).
It is closely related to Invariant 2a in that it expresses a similar state of the variables, but shifted forward by one row and column.

We will see that for both Invariants~2a and~2b  the resulting algorithms  shift some computation encountered in the algorithm that arises from Invariant~1 and fuses this into an adjacent iteration, thus simplifying the update of most of the trailing matrix.

\subsubsection
{Invariant~3 (left-looking variant)}
    {
    \footnotesize
\setlength{\arraycolsep}{2pt}
  \begin{eqnarray}
  \nonumber
\left( \begin{array}{c I c | c}
X_{TL} & \star & \star \\ \whline
x_{ML}^T & 0 & \star \\ \hline
X_{BL} & x_{BM} & X_{BR}
\end{array} \right)
=
\left( \begin{array}{c I c | c}
T_{TL} & \star & \star \\ \whline
\tau_{ML} e_l^T & 0 &  \star \\ \hline
0   & \widehat x_{BM}  & \widehat X_{BR}
\end{array} \right) \wedge 
\mbox{
(\ref{eqn:inv_unb_right_2})}
\comment{ 
\left( \begin{array}{c I c | c}
\widehat X_{TL} & 
\star  & 
\star \\ \whline
\widehat x_{ML}^T & 0 & 
\star \\ \hline
\widehat X_{BL} & \widehat x_{BM} & \widehat X_{BR}
\end{array} \right)
\\
\label{eqn:inv_unb_left-2}
& & ~~~~~ =
\left( \begin{array}{c I c | c}
\widetilde{L}_{TL} & 0 & 0 \\ \whline
\widetilde{l}_{ML}^T & 1 & 0 \\ \hline
\widetilde{L}_{BL} & \widetilde{l}_{BM} &  L_{BR}
\end{array}\right)
\left( \begin{array}{c I c | c}
T_{TL} & \star  & \star \\ \whline
\tau_{ML} e_l^T  & 0 & \star \\ \hline
0 & \tau_{BM} e_f & T_{BR}
\end{array} \right)
\left( \begin{array}{c I c | c}
\widetilde{L}_{TL}^T & \widetilde{l}_{ML} & \widetilde{L}_{BL}^T \\ \whline
0 & 1 & \widetilde{l}_{BM}^T \\ \hline
0 & 0 &  L_{BR}^T
\end{array}\right)
} 
.
\end{eqnarray}
    }


\subsubsection{Other invariants}

For operations like Cholesky factorization, there are also  ``bordered algorithms'' that have invariants where only the top-left submatrix has been computed with and updated with its final result.
Such invariants and corresponding variants also exist for our operation.  We choose not to pursue these since we are mostly interested in algorithms to which pivoting can be added.

\subsection{Unblocked algorithms} \label{sec:unblocked}

We now discuss unblocked algorithms that result from the systematic application of the FLAME methodology.  
Details of the derivation 
can be found in a companion technical report~\cite{vandegeijn2023derivingalgorithmstriangulartridiagonalization}.

\subsubsection{Right-looking (modified Parlett-Reid) algorithm}
If one chooses to maintain Invariant~1 and in each step a single new row and column are exposed, the right-looking algorithm given in Fig.~\ref{fig:LTLt_unb} (left) results.  

\paragraph*{Cost}
The dominant cost term of this  algorithm comes from the skew-symmetric rank-2 update. 
 If $ X $ is $ m \times m $ and $ X_{TL} $ is $ k \times k $, then $ X_{BR} $ is $ (m-k-1) \times (m-k-1) $ and updating it requires
approximately $ 2 (m-k-1) \times (m-k-1) $ flops (updating only the lower-triangular part).
The approximate total cost is hence
$
2 m^3/3 \mbox{~flops}$.

\paragraph*{Notes} This is the    Parlett-Reid algorithm for computing the $ L T L^T $ factorization of a symmetric matrix~\cite{ParlettReid},  modified for the skew-symmetric case in~\cite{Wimmer2012}.

\NoShow{
\subsection{Lazy right-looking algorithm}
\label{sec:unb-lazy}

A slight variation on the invariant
in (\ref{sec:unb-right})
that yields the right-looking algorithm is given by
{
 \setlength{\arraycolsep}{2pt}
\[
    \begin{array}{l}
\left( \begin{array}{c I c | c}
X_{TL} & \star & \star \\ \whline
x_{ML}^T & \chi_{MM} & \star \\ \hline
X_{BL} & x_{BM} & X_{BR}
\end{array} \right)
=
\left( \begin{array}{c I c | c}
T_{TL} & ~~~~~~~\star ~~~~~~~& \star \\ \whline
 \tau_{ML} e_l^T & 
 \multicolumn{2}{c}{
 \multirow{2}{*}{
 $
 \left( \begin{array}{c | c}
 1 &  0 \\ \hline
l_{BM} & L_{BR} 
\end{array}
\right)
 \left( \begin{array}{c | c}
 0 &  \star \\ \hline
\tau_{BM} 
e_f
& T_{BR} 
\end{array}
\right)
\left( \begin{array}{c | c}
 1 &  0 \\ \hline
l_{BM} & L_{BR} 
\end{array}
\right)^T
$
 }
 }
  \\ \cline{1-1}
0  &  \multicolumn{2}{c}{}
\end{array} \right) \wedge \\
~~ 
\color{gray}
\left( \begin{array}{c I c | c}
\widehat X_{TL} & \star & \star \\ \whline
\widehat x_{ML}^T & 0 & \star \\ \hline
\widehat X_{BL} & \widehat x_{BM} & \widehat X_{BR}
\end{array} \right) 
 =
\left( \begin{array}{c I c | c}
\widetilde{L}_{TL} & 0 & 0 \\ \whline
\widetilde{l}_{ML}^T & 1 & 0 \\ \hline
\widetilde{L}_{BL} & \widetilde{l}_{BM} &  I
\end{array}\right)
\left( \begin{array}{c I c | c}
T_{TL} & - \tau_{ML} e_l  & 0 \\ \whline
\tau_{ML} e_l^T  & 0 & -\tau_{BM} (L_{BR} e_f)^T \\ \hline
0 & \tau_{BM} L_{BR} e_f & L_{BR} T_{BR} L_{BR}^T
\end{array} \right)
\left( \begin{array}{c I c | c}
\widetilde{L}_{TL}^T & \widetilde{l}_{ML} & \widetilde{L}_{BL}^T \\ \whline
0 & 1 & \widetilde{l}_{BM}^T \\ \hline
0 & 0 &  I
\end{array}\right).
\end{array}
    \]
}%
This invariant captures that the latest Gauss transform, defined by $ l_{BM} $ has been computed, but not yet applied to the remainder of $ X $.
Going through the motions of derivation yields the  algorithm also given in Fig.~\ref{fig:LTLt_unb_right}.
}  

\subsubsection{Two-step  right-looking  algorithm}

\label{sec:unb-right-wimmer}

\begin{figure}[tbp]
    \input LTLt_unb_2_step
    
\centering
    \small
    \FlaAlgorithm    
    \caption{Two-step right-looking unblocked  algorithm. \NoShow{It is the addition of the commands in red (along with the ensuing changes to updates of $X$ in blue) that distinguishes our algorithm for computing all of $ L $ from Wimmer's algorithm for computing only the Pfaffian of $ X $.}}
    \label{fig:LTLt_unb_Wimmer}
\end{figure}

In Fig.~\ref{fig:LTLt_unb} (left), observe that because of the special structure of skew-symmetric matrices,  in the right-looking (modified Parlett-Reid) algorithm the application of the current Gauss transform does not change the ``next column,'' 
$
x_{32} $, since $ \chi_{22} = 0 $.
This suggests exposing two columns in a given iteration.
Starting with Invariant~1, the FLAME methodology yields what we call the two-step right-looking algorithm given in Fig.~\ref{fig:LTLt_unb_Wimmer}.  See~\cite{vandegeijn2023derivingalgorithmstriangulartridiagonalization} for details.

\paragraph*{Cost}
It is in the skew-symmetric rank-2 update of $ X_{44} $  that most of the operations are performed, yielding an approximate cost 
of
$
m^3/3 \mbox{~flops}$,
or {\bf half} of the cost of the more straight-forward unblocked right-looking 
(modified Parlett-Reid) algorithm.

\paragraph*{Notes}
Our algorithm is essentially Wimmer's algorithm in~\cite{Wimmer2012}.  However, his algorithm skips the computation of $ l_{43} $ (which defines the second Gauss transform in a two-step iteration) and $\tau_{32}$,
since only every other subdiagonal element of the tridiagonal matrix was required for the computation of the Pfaffian. 
Later, the connection between our two-step algorithm and Wimmer's algorithm can be seen through the blocked version of Variant 1 (Fig.~\ref{fig:LTLt_blk}), if a block size of two is employed. 
This is equivalent to the two-step algorithm, with some rearrangement of the computation. By noting that the block {\sc skew\_tridiag\_rankk} operation is a single {\sc skew\_rank2} update and then deleting the second {\sc skew\_rank2} update and the computation of the corresponding column of $L$, one arrives back at Wimmer's algorithm.
Wimmer's implementation (PFAPACK) reverts back to the unblocked right-looking (Parlett-Reid) algorithm when the full $ L T L^T $ output is required.
Thus, our algorithm ``completes'' Wimmer's work and is beneficial for situations where the full $ LTL^T $ factorization is needed, for example when fast-updating computed Pfaffians~\cite{blockedvmc} or solving $ X v = w $.

\NoShow{
As of this writing, we have not attempted to derive a  two-step algorithm for \LTLt.  We suspect  that the zeroes on  the diagonal  of a skew-symmetric matrix are key to Wimmer's algorithm for \skewLTLt\ and that hence there is no  beneficial equivalent algorithm for \LTLt.
}

\subsubsection{Left-looking (modified Aasen's) algorithm}

\label{sec:unb-left}

Maintaining Invariant~3 and exposing in each step a single new column yields the left-looking algorithm given in Fig.~\ref{fig:LTLt_unb} (right).
That algorithm works whether the first column of $ L $ is $ e_0 $ or not.

\paragraph*{Cost}
The dominant cost term of this algorithm comes from  
updating $\renewcommand{\arraystretch}{1.0} \left( \begin{array}{c}
\chi_{21} \\ \hline
x_{31}
\end{array}
\right) $.
Since $ \renewcommand{\arraystretch}{1.2}\left( \begin{array}{c|c}
T_{00} & \star \\ \hline
\tau_{10} e_l^T & 0
\end{array}
\right) $ (now stored in $ \renewcommand{\arraystretch}{1.2}\left( \begin{array}{c|c}
X_{00} & \star \\ \hline
x_{10}^T & 0
\end{array}
\right) $)  is skew-symmetric and tridiagonal, this incurs roughly\footnote{
Not including a lower order term which may be affected by whether the first column equals zero or not.} the cost of a matrix-vector multiplication with matrix
\textcolor{review}{$\renewcommand{\arraystretch}{1.0}\left( \begin{array}{c | c  }
l_{20}^T & \lambda_{21}   \\ \hline
L_{30} & l_{31}  
\end{array} \right)$}, or approximately
$ 2 (m-k) \times k $ flops if $ X_{TL} $ is $ k \times k $.
The approximate total cost is hence roughly $\sum_{k=0}^{m-2}
2 k( m-k )  = m^3/3 $ flops.
\NoShow{
\begin{equation}
    \nonumber
\sum_{k=0}^{m-2}
2 k( m-k ) =
2 \left( \sum_{k=0}^{m-2}
k m 
-
\sum_{k=0}^{m-2}k^2  \right)
\approx
2 \left(
m^3/2 - m^3/3 \right) = m^3 / 3  \mbox{~flops}.
\end{equation}
}
This is {\bf half} the approximate cost of the unblocked right-looking (Parlett-Reid) algorithm and matches the approximate cost of the two-step right-looking (Wimmer's) algorithm.

\paragraph*{Notes} This algorithm is a twist on Aasen's algorithm~\cite{Aasen}.
Aasen recognized that
$ X = L T L^T = L H $, where $ H = T L^T $ is upper-Hessenberg.  As noted in his paper, in each iteration only one column of $H $needs to be computed and used in an iteration and  hence $ H $ needs not be stored.  This column of $ H $ is  
$
\left( \begin{array}{c | c  }
X_{00} & 
\star   \\ \hline
x_{10}^T & 0
\end{array}
\right)
\left( \begin{array}{c}
l_{10}  \\ \hline
1  
\end{array} \right)
$
in our algorithm.
The difference in number of flops performed between right-looking (Parlett-Reid) and left-looking (Aasen) algorithms was noted by Wimmer, and is one motivation for computing a ``partial'' right-looking factorization (useful for computing the Pfaffian).
Whether one can derive a left-looking partial factorization algorithm which performs even fewer operations is an open question.

\subsubsection{New level-2 BLAS-like operations}

 \begin{figure}[tb!]
 \begin{center}
 \setlength\tabcolsep{4pt}
 \begin{tabular}{|l|l|l|}
  \hline 
     {\bf Operation} & {\bf Name} & {\bf Used by} \\ \whline
     \multicolumn{3}{|l|}{
     {\bf Level-2 BLAS-like}} \\ \whline
     $ A := \beta A + \alpha ( x y^T - y x^T ) $ & {\sc skew\_rank2}
     & unblocked right-looking 
     \\ \hline
          $ A := \textcolor{review}{\gamma A + \alpha x u^T + \beta y v^T} $ & {\sc gen\_rank2}
     & unblocked right-looking 
     \\ 
     & &   unblocked 2-step \\
     & &   right-looking  \\ \hline
     $ y := \beta y + \alpha A T x $ & {\sc skew\_tridiag\_gemv} &
     unblocked left-looking  \\ \whline
\multicolumn{3}{|l|}{
     {\bf Level-3 BLAS-like}} \\ \whline
     $ C := \beta C + \alpha A T A^T $ & {\sc skew\_tridiag\_rankk}
     & blocked right-looking, 
     \\ 
     & &  fused blocked right-looking  \\ \hline
     $ C := \beta C +\alpha A T B $ & {\sc skew\_tridiag\_gemm} &
     blocked left-looking \\ \hline
     $ C := \beta C +\alpha( A B^T  - B A^T )  $ &
     {\sc skew\_rank2k} & blocked Wimmer's    \\ 
     & & (Section~\ref{sec:blk-right-wimmer}, first option) \\ \hline
\end{tabular}
 \end{center}
 \caption{BLAS-like operations encountered in the various algorithms for the $ L T L^T $ factorization.}
 \label{fig:BLAS3-like}
 \label{fig:BLAS2-like}
 \end{figure}

 Fig.~\ref{fig:BLAS2-like} gives new level-2 BLAS-like operations encountered in the derived algorithms. \textcolor{review2}{The BLAS interface does not traditionally incorporate any skew-symmetric operations, although recently several new operations (some of which appear in Fig.~\ref{fig:BLAS2-like}) have been added to the reference implementation \cite{pr1,pr2}.}
{\sc gen\_rank2} is included since in Section~\ref{sec:blk} we find that the right-looking unblocked algorithm may be used to factor only the current panel of columns.  The encountered {\sc skew\_rank2} only updates that panel so that either a rather complicated level-2 BLAS-like operation needs to be defined to operate on trapezoidal regions, or the operation can be split into a square {\sc skew\_rank2} and a rectangular {\sc gen\_rank2} update,  which is what we chose to do.

\subsection{Blocked algorithms}
\label{sec:blk}

High performance for DLA operations  can be attained by casting them in terms of matrix-matrix operations (level-3 BLAS)~\cite{BLAS3}.  We now discuss such blocked algorithms, explosing new level-3 BLAS-like operations.

\subsubsection{Right-looking algorithm}
\label{sec:blk-right}

Starting with Invariant~1, a blocked algorithm can be derived by repartitioning the matrix $ X $ as
{
\small
\setlength{\unitlength}{2pt}
\[
\left( \begin{array}{c I c | c}
X_{TL} & \star & \star \\ \whline
x_{ML}^T & 0 & \star \\ \hline
X_{BL} & x_{BM} & X_{BR}
\end{array} \right) \rightarrow \left( \begin{array}{c I c | c | c | c }
X_{00} & \cellcolor{lightgray!35}  \star & \cellcolor{lightgray!35}  \star & \star & \star \\ \whline
\cellcolor{lightgray!35} x_{10}^T & \cellcolor{lightgray!35} 0 & 
\cellcolor{lightgray!35} \star & \cellcolor{lightgray!35} \star& \cellcolor{lightgray!35} \star \\ \hline
\cellcolor{lightgray!35} X_{20} & 
\cellcolor{lightgray!35} x _{21} & \cellcolor{lightgray!35} X_{22} &  \cellcolor{lightgray!35} \star & \cellcolor{lightgray!35} \star \\ \hline
x_{30}^T & \cellcolor{lightgray!35} \chi_{31}
& \cellcolor{lightgray!35} x_{32}^T & 0
& \star \\ \hline
X_{40} & \cellcolor{lightgray!35} x_{41}
& \cellcolor{lightgray!35} X_{42} & x_{43} & X_{44}
\end{array} \right),
\]
}%
where the gray highlighting captures the block of rows and columns which will be factorized in this iteration.  The other matrices are repartitioned conformally.  The FLAME methodology yields the algorithm in Fig.~\ref{fig:LTLt_blk} (Variant~1).
Importantly, this algorithm
casts the bulk of the computation in terms of the 
``sandwiched'' skew-symmetric rank-k update ({\sc skew\_tridiag\_rankk}) of $ \renewcommand{\arraystretch}{1.0}
\left( \begin{array}{c | c}
0 & \star \\ \hline
x_{43} & X_{44} \end{array} \right)$.

\paragraph*{Cost}
What is somewhat surprising about this blocked right-looking algorithm  is that its cost, when the blocking size $ b $ is reasonably large, is essentially 
$ m^3 / 3 $ flops which equals half the cost of the unblocked right-looking (modified Parlett-Reid) algorithm and is equal to that of the unblocked left-looking (modified Aasen) algorithm. \textcolor{review}{ As shown in Sec.~\ref{sec:skew}, the {\sc skew\_tridiag\_rankk} operation is equivalent to a series of rank-2 updates, which is the form of the straightforward blocked Parlett-Reid algorithm. However, blocked Parlett-Reid does not lead to the alternating zero column structure responsible for the reduction in cost of blocked Variant 1.}

\begin{figure}[tb!]

\resetsteps      


\renewcommand{\routinename}{ \left[ X, L \right] := \mbox{\sc LTLt\_blk}( X ) }


\renewcommand{\guard}{
  m( X_{TL} ) < m( X )-1
}


\renewcommand{\partitionings}{
$ L  = I $\\
  $
  X \rightarrow
  \left( \begin{array}{c I c | c}
  X_{TL} & \star & \star \\ \whline
  x_{ML}^T & 0 & \star \\ \hline
  X_{BL} & x_{BM} & X_{BR}
  \end{array}
  \right)
  $
,
  $
  L \rightarrow
\left( \begin{array}{c I c | c}
  L_{TL} & 0 & 0 \\ \whline
  l_{ML}^T & 1 & 0 \\ \hline
  L_{BL} & l_{BM} & L_{BR}
  \end{array}
  \right)
  $ 
}

\renewcommand{\moreinitialize}{
 \\
 \mbox{{\color{blue} if} Variant 2b Compute first Gauss transform:} \\
  \quad
  \begin{tabular}{@{}l @{\quad} l}
  $\chi = $ first element of $x_{BM}$ \\
  $ L_{BR} e_f := x_{BM} / \chi $ &
  \mbox{(compute first column of $L_{BR}$)} \\
  $ x_{BM} := \chi e_f $ \\
  \end{tabular} \\
  \mbox{\color{blue} endif}}

\renewcommand{\partitionsizes}{
$ X_{TL} $ and $ L_{TL} $ are $ 0 \times 0 $
}


\renewcommand{\repartitionings}{
\setlength{\arraycolsep}{2pt}
\footnotesize
$  \left( \begin{array}{c I c | c}
  X_{TL} & \star & \star \\ \whline
  x_{ML}^T & 0 & \star \\ \hline
  X_{BL} & x_{BM} & X_{BR}
  \end{array}
  \right)
  \rightarrow
  \left( \begin{array}{c I c | c | c | c}
  X_{00} & \star & \star & \star & \star \\ \whline
   x_{10}^T & 0 & \star & \star & \star\\ \hline
   X_{20} & x_{21} & X_{22} & \star & \star \\ \hline
   x_{30}^T & \chi_{31} & x_{32}^T & 0 & \star \\ \hline
   X_{40} & x_{41} & X_{42} & x_{43} & X_{44}
   \end{array} \right)
   $,
   $
   \left( \begin{array}{c I c | c}
  L_{TL} & 0 & 0 \\ \whline
  l_{ML}^T & 1 & 0 \\ \hline
  L_{BL} & l_{BM} & L_{BR}
  \end{array}
  \right)
  \rightarrow
  \cdots
   $ \vspace{3pt}
   }

\renewcommand{\repartitionsizes}{
  $ \chi_{ij} $ and $ \lambda_{ij}$ are scalars ...
  }


\renewcommand{\moveboundaries}{
\setlength{\arraycolsep}{2pt}
\footnotesize
\vspace{-5pt}
$  \left( \begin{array}{c I c | c}
  X_{TL} & \star & \star \\ \whline
  x_{ML}^T & 0 & \star \\ \hline
  X_{BL} & x_{BM} & X_{BR}
  \end{array}
  \right)
  \leftarrow
  \left( \begin{array}{c | c | c I c | c}
  X_{00} & \star & \star & \star & \star \\ \hline
   x_{10}^T & 0 & \star & \star & \star \\ \hline
   X_{20} & x_{21} & X_{22} & \star & \star \\ \whline
   x_{30}^T & \chi_{31} & x_{32}^T & 0 & \star \\ \hline
   X_{40} & x_{41} & X_{42} & x_{43} & X_{44}
   \end{array} \right)
   $,
   $
   \left( \begin{array}{c I c | c}
  L_{TL} & 0 & 0 \\ \whline
  l_{ML}^T & 1 & 0 \\ \hline
  L_{BL} & l_{BM} & L_{BR}
  \end{array}
  \right)
  \leftarrow
  \cdots
   $}


\renewcommand{\update}{
\vspace{-5pt}
\setlength{\arraycolsep}{2.5pt}
$
  \begin{array}{ l } 
  \mbox{\underline{\bf Right-looking (Variant~1):}} \vspace{3pt} \\
  \begin{array}[t]{@{}l@{}}
    \begin{array}{@{}l@{}}
    \left[ \left( \begin{array}{c | c}
    0 & \star \\ \hline
    x_{21} & X_{22} \\ \hline
    \chi_{31} & x_{32}^T \\ \hline
    x_{41} & X_{42}
    \end{array} \right),
    \left( \begin{array}{c | c | c }
1 & 0 & 0 \\ \hline
l_{21} & L_{22} & 0   \\ \hline
\lambda_{31} & l_{32}^T &
1 \\ \hline
l_{41} & L_{42} &
l_{43} 
\end{array}
\right)
    \right]   :=
    \mbox{\sc LTLt\_unb\_0}( 
    \left(
    \begin{array}{c | c}
    0 & \star \\ \hline
    x_{21} & X_{22} \\ \hline
    \chi_{31} & x_{32}^T \\ \hline
    x_{41} & X_{42}
    \end{array} \right), \left( \begin{array}{c | c | c }
1 & 0 & 0 \\ \hline
l_{21} & L_{22} & 0   \\ \hline
\lambda_{31} & l_{32}^T &
1 \\ \hline
l_{41} & L_{42} &
l_{43} 
\end{array}
\right)
    )   
    \end{array}
    \\  
        \left( \begin{array}{c | c}
    0 & \star \\ \hline
    x_{43}& X_{44}
    \end{array}
    \right)
    := 
        \left( \begin{array}{c | c}
    0 & \star \\ \hline
    x_{43}& X_{44}
    \end{array}
    \right) - \left( \begin{array}{ c | c }
l_{32}^T &
1 \\ \hline
L_{42} &
l_{43} 
\end{array}
\right)
\left( \begin{array}{c | c }
 X_{22} & \star  \\ \hline
 x_{32}^T &
0 
\end{array}
\right)
\left( \begin{array}{c | c}
l_{32}  &  L_{42}^T \\ \hline
1 & l_{43}^T 
\end{array}
\right) \\
X_{44} := X_{44} +  
( l_{43} x_{43}^T - x_{43} l_{43}^T )
    \end{array}
  \\
  \mbox{\underline{\bf Left-looking (Variant~3):}} \vspace{3pt} \\
    \begin{array}{@{}l}
    \begin{array}{@{}l}
    \left( \begin{array}{ c | c } 
x _{21} & X_{22} \\ \hline
\chi_{31}
&  x_{32}^T \\ \hline
x_{41}
&  X_{42} 
\end{array} \right) :=
\left( \begin{array}{ c | c } 
x _{21} & X_{22} \\ \hline
\chi_{31}
&  x_{32}^T \\ \hline
x_{41}
&  X_{42} 
\end{array} \right) -
\left( \begin{array}{c I c }
 L_{20} & 
 l _{21}  \\ \hline
 l_{30}^T &  \lambda_{31}
\\ \hline
 L_{40} &  l_{41}
\end{array} \right)
\left( \begin{array}{c I c  }
 X_{00} &  \star  \\ \whline
 x_{10}^T & 0 
\end{array} \right)
\left( \begin{array}{c | c }
 l_{10}  &  L_{20}^T  \\ \whline
1 & 
 l_{21}^T 
\end{array} \right)
\\
    \left[ \left( \begin{array}{c | c}
    0 & \star \\ \hline
    x_{21} & X_{22} \\ \hline
    \chi_{31} & x_{32}^T \\ \hline
    x_{41} & X_{42}
    \end{array} \right),
    \left( \begin{array}{c | c | c }
1 & 0 & 0 \\ \hline
l_{21} & L_{22} & 0   \\ \hline
\lambda_{31} & l_{32}^T &
1 \\ \hline
l_{41} & L_{42} &
l_{43} 
\end{array}
\right)
    \right]
 :=
    \mbox{\sc LTLt\_unb}( 
    \left(
    \begin{array}{c | c}
    0 & \star \\ \hline
    x_{21} & X_{22} \\ \hline
    \chi_{31} & x_{32}^T \\ \hline
    x_{41} & X_{42}
    \end{array} \right),
         \left( \begin{array}{c | c | c }
1 & 0 & 0 \\ \hline
l_{21} & L_{22} & 0   \\ \hline
\lambda_{31} & l_{32}^T &
1 \\ \hline
l_{41} & L_{42} &
l_{43} 
\end{array}
\right)
    )   \vspace{3pt}
    \end{array}
        \end{array}
\end{array}
$
}

\centering
\footnotesize
    \FlaAlgorithm    
    \caption{\textcolor{review}{Blocked algorithms.  In calling the unblocked algorithm {\sc ...\_unb} indicates the first column of $ L $ is to be used as passed in while {\sc ...\_unb\_0} means that that first column is implicitly equal to the first standard basis vector.}}
    \label{fig:LTLt_blk}
\end{figure}

\begin{figure}

\centering
\footnotesize
     \input LTLt_blk_fused
    \caption{\textcolor{review}{Fused blocked algorithms.}}
    \label{fig:LTLt_blk_fused}
\end{figure}

\paragraph*{Notes} If we choose the block size in the  algorithm equal to one ($ X_{22} $ is $ 0 \times 0  $), then this becomes the 
unblocked right-looking (Parlett-Reid) algorithm.  
This new  blocked algorithm has some resemblance  to the blocked algorithm for computing the $ L T L^T $ factorization of a symmetric matrix by Rozlo\v{z}n\'{\i}k et al.~\cite{Miroslav2011} and can be modified to perform that operation. 
In their algorithm, blocks of the same matrix $ H $ that Aasen introduced are computed, which is what we avoid.  {\em If} a high-performance sandwiched {\sc skew\_tridiag\_rankk} update were available, {\em then} our algorithm  avoids the workspace required  by their algorithm for parts of $ H $, further discussed in Section~\ref{sec:implementation}.

\subsubsection{Factoring the panel}

The factoring of the panel in the various blocked algorithms is accomplished by calling any of the unblocked algorithms, provided they are modified to not update any part of the matrix $ X $ outside the current panel.  

\subsubsection{Fused right-looking algorithms}

A problem with the algorithm in Section~\ref{sec:blk-right} is that the separate {\sc skew\_rank2}  operation, $X_{44} := X_{44} + ( l_{43} x_{43}^T - x_{43} l_{43}^T) $,
requires additional memory accesses
for an extra pass over  $ X_{44} $.
The solution is to delay the application of the last Gauss transform computed during the factorization of the current panel, thus fusing it with the update in the next iteration.
This is what Invariant~2a prescribes.
Alternatively, in the current iteration, one can  compute an additional Gauss transform in the current panel factorization, but not yet apply it, as prescribed by Invariant~2b.  This also fuses the application of that Gauss transform with the next iteration, but shifts what data is accessed by one column relative to Invariant~2a.  
The resulting blocked variants are shown in Fig.~\ref{fig:LTLt_blk}.

\paragraph*{Cost}
The cost of these, in flops, is essentially that of blocked  Variant~1. 

\paragraph*{Notes}
The benefit of these algorithms is that the update is \textcolor{review}{recast as a single level-3 BLAS-like} sandwiched skew-symmetric rank-k update ({\sc skew\_tridiag\_rankk})\textcolor{review}{, better reflecting the underlying structure of the problem and} making it  higher performing \textcolor{review}{via a fused implementation}, as we will see in Section~\ref{sec:results}.

\subsubsection{Left-looking algorithm}

Invariant~3 can  be used to derive the blocked left-looking algorithm given in Fig.~\ref{fig:LTLt_blk}.

\paragraph*{Cost} The cost of this algorithm is essentially the same as the unblocked left-looking algorithm and the other blocked algorithms that we have discussed.

\paragraph*{Notes}
A blocked variant of Aasen's (left-looking) algorithm for the symmetric case was derived by Ballard et al~\cite{ballard_implementing_2013} and implemented using tile-based parallelism.

\subsubsection{Two-step blocked algorithm}

\label{sec:blk-right-wimmer}

There are a number of ways to arrive at blocked algorithms that are similar to the two step (Wimmer's) unblocked  algorithm in~\cite{Wimmer2012}, all of which appear to be new.  Here we discuss a few:

\paragraph*{Deriving  Wimmer's blocked algorithm}

Wimmer's blocked algorithm can be derived by starting the derivation process with  the postcondition of the algorithm as $ X = T \wedge \widehat X = W L^T - L W^T \wedge W = L S \wedge T = S - S^T $, with implicit assumptions about the structure of the various operands and an understanding that in the end only $ X $ and $ L $ are returned as results.
Only the part of $ W $ needed in the current iteration needs to be kept.
This approach may end up yielding more variants.

\paragraph*{Accumulating a rank-2k update}

The update 
$ X_{44} := l_{43} x_{43}^T - x_{43} l_{43}^T $ outside of the panel factorization can be delayed until the completion of that factorization.
This means accumulating the appropriate vectors as columns in matrices so that in the end a skew-symmetric rank-2k update can be performed.

\paragraph*{Forming $ W$ after the panel factorization}
Alternatively, we can transform the 
$ C:=C - A T A^T$ update with $ T $ of size $ b \times b $  of the right-looking algorithm in Fig.~\ref{fig:LTLt_blk} into a skew-symmetric rank-2k update with $ k=\lfloor b/2 \rfloor $ by invoking Theorem~\ref{thm:TandS}: $ W = A S $ followed by $ C := C-W A^T - A W^T $, where every other column of $ W $ equals zero, starting with the first column. For the fused right-looking algorithms \textcolor{review}{(Fig.~\ref{fig:LTLt_blk_fused})}, the 
$ C:=C - A T A^T$ update with $ T $ of size $ (b+1) \times (b+1) $  of the fused right-looking algorithms is transformed into a skew-symmetric rank-2k update with $ k=\lfloor b/2 \rfloor$, reducing the computational cost to that of the two-step unblocked algorithm\textcolor{review}{.}
In other words, the two-step algorithm  is attained via  an implementation detail for this level-3 BLAS-like operation. 

\paragraph*{Cost}
The cost is asymptotically that of the other blocked  algorithms.  

\paragraph*{Note}
Wimmer accumulates rank-2k updates in his paper~\cite{Wimmer2012}, yielding a blocked algorithm with a cost of \textcolor{review}{roughly} $m^3/3$ flops {\em when only the Pfaffian is required}.  When the full factorization is required, he reverts back to the unblocked right-looking (modified \textcolor{review}{Parlett}-Reid) algorithm and accumulates those rank-2 updates, yielding an algorithm that performs roughly $ 2 m^3 /3 $ flops.
Our algorithms  compute all of $ L $.

\subsubsection{New level-3 BLAS-like operations}

In Fig.~\ref{fig:BLAS3-like}, we give new level-3 BLAS-like operations that show up in the derived algorithms.

\subsection{Pivoted algorithms}\label{sec:pivoting}

As for LU factorization, pivoting can be added to improve numerical stability. While in~\cite{vandegeijn2023derivingalgorithmstriangulartridiagonalization} we discuss how such algorithms can be systematically derived, here we give an intuitive discussion.

\subsubsection{Pivot matrices}

We start by establishing some notation.

\begin{definition}
Given vector $ x $, 
$
\mbox{\sc iamax}(x)
$
returns  the index of the element in $ x $ with largest magnitude.
(In our discussion, indexing starts at zero).
\end{definition}

\begin{definition}
\label{def:perm_vector}
Given nonnegative integer $ \pi $, the matrix $ P( \pi ) $ is the permutation matrix of appropriate size that, when applied to a vector $ x $, swaps the top element, $ \chi_0 $, with the element indexed by $ \pi $, $ \chi_{\pi} $.
\end{definition}
Applying $ P( \pi ) $ to  $ m \times n $ matrix $ A $ from the left swaps the top row with the row indexed with $ \pi $.  From the context we know that in this case $ P( \pi ) $ is $ m \times m $.  
We will accumulate $k \le m$ pivots into a {\em permutation vector} 
$ p = ( \pi_0, \pi_1, \ldots , \pi_{k-1} )^T $
\NoShow{$ p = \left( \begin{array}{c}
\pi_0 \\
\vdots \\
\pi_{k-1} 
\end{array}
\right) $}
that describes the combined permutation matrix
\[
P( p ) = 
\left( \begin{array}{c | c}
I_{(k-1) \times (k-1)} & 0 \\ \hline
0 & P( \pi_{k-1} )
\end{array}
\right)
\cdots 
\left( \begin{array}{c | c}
1 & 0 \\ \hline
0 & P( \pi_1 )
\end{array}
\right)
P( \pi_0 ),
\]
where $ I_{l \times l} $ is an $ l \times l $ identity matrix.

\subsubsection{Adding pivoting to unblocked algorithms}

\NoShow{
\begin{figure}[tbp]
    \input LTLt_piv_unb_right_left.tex
    
\centering
{
    \footnotesize
    \FlaAlgorithm 
    }
    \caption{Pivoted unblocked  algorithm.}
    \label{fig:LTLt_piv_unb_right_left}
\end{figure}
}

Adding pivoting to the 
unblocked algorithms as indicated in red in Fig.~\ref{fig:LTLt_unb} is  straightforward: before a new Gauss transform is computed, the largest value in magnitude in the vector 
$ \renewcommand{\arraystretch}{1.0} \left( \begin{array}{c}
\chi_{21} \\ \hline
x_{31}
\end{array}
\right)
$ is found and swapped to the top, 
swapping those rows in the part of $ L $ that was previously computed, as well as rows and columns in 
$ \renewcommand{\arraystretch}{1.0} \left( \begin{array}{c | c}
0 & \star\\ \hline
x_{32} & X_{33}
\end{array}
\right)
$.  

\subsubsection{Adding pivoting to blocked algorithms}

\begin{figure}[tbp]

\centering
{
    \footnotesize
    \input LTLt_piv_blk.tex
    }
    \caption{Pivoted blocked right-looking algorithms. All of the remaining parts of $ X$ are passed into the panel factorization so that symmetric pivoting can be applied.  The matrix $X$ is only updated (other than pivoting) up to the double lines. The vector $p_2^\star$ omits the first element, while the scalar $p_4^f$ is the first element of $p_4$ only. The very first pivot, $\pi_0$, is not computed and is assumed to be zero. }
    \label{fig:LTLt_piv_blk}
\end{figure}

Pivoting can only be added to right-looking blocked algorithms. 
A blocked pivoted left-looking algorithm cannot exist, since, in order to update the current panel, which columns to pivot into that panel  would need to be known. 
However, some of those columns are determined during the factorization of earlier columns in that panel and so cannot be pre-selected.

For the panel factorization within  blocked right-looking algorithms, an unblocked left-looking algorithm (with pivoting)  must be used.
The panel factorization must employ a left-looking pivoted algorithm because the pivoting of that panel will potentially bring in columns from the trailing part of $ X $ that has not yet been updated. Thus, the invariant must reflect that the next column to factorize has not yet had updates applied from the current panel, which contradicts the right-looking invariant.

During the panel factorization, columns may be pivoted in (based on the position of the maximum element in $\renewcommand{\arraystretch}{1.0} \left( \begin{array}{c} 
\chi_{21} \\ \hline
x_{31}
\end{array}
\right)
$) from either within the panel or from the trailing portion of $ X $. Thus, the entire unfactorized portion of $ X $ is passed into the panel factorization, although only part of it will be factorized and updated.
Resulting pivoted blocked right-looking algorithms are given in Fig.~\ref{fig:LTLt_piv_blk}.
The other blocked right-looking algorithms can be similarly modified.

\section{Implementation}\label{sec:implementation}

We now turn to translating theory into practice.

\subsection{Storage details}\label{sec:storage}

It is desirable to reuse the storage of the input matrix for the decomposed form.  
\NoShow{For example, the LU factorization as implemented in LAPACK~\cite{LAPACK3} overwrites the input matrix $ A $ with the factors $ L $ and $ U $, where the unit diagonally elements of $ L $ are not stored.  We now discuss how to achieve this having derived the various algorithms%
\footnote{Alternatively, one could specify this in the pre- and postconditions, and derive algorithms with these assumptions.}.}
For an $X=LTL^T$ decomposition, the strictly lower (or upper)  elements of $ X $ can hold the unique non-unit and non-zero elements of $T$ and $L$. 
In particular, the lower part of $L$ (except the leading and presumably zero column) may be stored, ``shifted'' left by one column, below the sub-diagonal of $X$, with elements of $T$ stored on the sub-diagonal since the diagonal elements of $ L $ are known to be ones.  Incorporating this  would mean no workspace is needed. 

The problem is that some updates  explicitly reference the diagonal of $L$, which is implicitly one. For example, in the blocked right-looking algorithm, the update
\begin{align}
\setlength\arraycolsep{2pt}
\left( \begin{array}{c | c}
    \chi_{33} & \star \\ \hline
    x_{43}& X_{44}
    \end{array}
    \right)
    &
    := 
    \left( \begin{array}{c | c}
    \chi_{33} & \star \\ \hline
    x_{43}& X_{44}
    \end{array}
    \right) - \left( \begin{array}{ c | c }
l_{32}^T &
1 \\ \hline
L_{42} &
l_{43} 
\end{array}
\right)
\left( \begin{array}{c | c }
 X_{22} & \star  \\ \hline
 x_{32}^T &
0 
\end{array}
\right)
\left( \begin{array}{c | c}
l_{32}  &  L_{42}^T \\ \hline
1 & l_{43}^T 
\end{array}
\right) \nonumber \\
\label{eqn:updateLTLt}
&\;=
\left( \begin{array}{c | c}
    \chi_{33} & \star \\ \hline
    x_{43}& X_{44}
    \end{array}
    \right) - \left( \begin{array}{ c | c }
l_{32}^T &
1 \\ \hline
L_{42} &
l_{43} 
\end{array}
\right)
\left( \begin{array}{c | c }
 T_{22} & \star  \\ \hline
 \tau_{32} e_l^T &
0 
\end{array}
\right)
\left( \begin{array}{c | c}
l_{32}  &  L_{42}^T \\ \hline
1 & l_{43}^T 
\end{array}
\right)
\end{align}
appears.
When stored as described above, the ``$1$'' 
\NoShow{which is in fact $\lambda_{33}$} refers to the same storage as the last element of $x_{32}^T $, which 
\NoShow{in turn} physically stores $\tau_{32}$. 
This value cannot be temporarily overwritten with an explicit 1, as it is also referenced in the central tridiagonal matrix in the update. One option is to split the update into separate steps, for example,
\begin{align}
\label{eqn:split_update1}
\setlength\arraycolsep{2pt}
x_{43} &:= x_{43} - (L_{42} T_{22} l_{32} + \tau_{32} l_{43} e_l^T l_{32} -  \tau_{32} L_{42} e_l) \\
\label{eqn:split_update2}
\setlength\arraycolsep{2pt}
X_{44} &:= X_{44} - \left( \begin{array}{ c | c }
L_{42} &
l_{43} 
\end{array}
\right)
\left( \begin{array}{c | c }
 T_{22} & \star  \\ \hline
 \tau_{32} e_l^T &
0 
\end{array}
\right)
\left( \begin{array}{c}
L_{42}^T \\ \hline
l_{43}^T 
\end{array}
\right).
\end{align}
This  requires $ L_{42} $ to be brought into memory twice (a nontrivial overhead.)  Instead, we  store the subdiagonal elements of $T$ in an external vector, and to store explicit 1's on the subdiagonal of $X$ as columns of $L$ are computed. This allows the entire update to be achieved using the {\sc skew\_tridiag\_rankk}  operation from Fig.~\ref{fig:BLAS3-like}.  Upon completion, the  elements of $ T $ can be moved to overwrite the first subdiagonal of $ X $.

\comment{Interestingly, the choice of the storage location of the $L$ and $T$ matrices is expressible as a part of the PME and invariants by imposing additional conditions. Dependency analysis of the resulting logical expressions at the ``before'' and ``after'' points can then illuminate when such fusion optimizations can and cannot be performed. Storing $T$ as an external array also allows elements to be accessed contiguously.}

\subsection{C++ interface}

\begin{figure}[tb!]

\footnotesize
\begin{center}
\begin{tabular}{l}
\begin{lstlisting}[language=C++]
void ltlt_pivot_blockRL_var0(const matrix_view<double>& X,
                             const    row_view<double>& t,
                             const    row_view<   int>& p,
                             int block_size)
{
    // Store L, with explicit '1's, below the diagonal of X
    // which requires a 1-column index shift
    matrix_view<double> L = X.rebased(1, 1);

    p[0] = 0;

    // Initialize T = [], m = 0, B = [1,n)
    auto [T, m, B] =
        partition_rows<DYNAMIC,1,DYNAMIC>(X);

    while (B)
    {
        // (  T ||  m |       B      )
        // ( R0 || r1 | R2 | r3 | R4 )
        auto [R0, r1, R2, r3, R4] =
            repartition<DYNAMIC,1>(T, m, B, block_size);

        ltlt_pivot_unblockLL(X[r1|R2|r3|R4][r1|R2|r3|R4],
                             t[r1|R2],
                             p[R2|r3],
                             false);

        pivot_rows(X[R2|r3|R4][R0],
                   p[R2|r3]);

        skew_tridiag_rankk('L',
                           -1.0, L    [r3|R4][R2|r3],
                                 t           [R2   ],
                            1.0, X    [r3|R4]       [r3|R4]);

        skew_rank2('L', 1.0, L[R4][r3],
                             X[R4][r3],
                        1.0, X[R4][R4]);

        // ( R0 | r1 | R2 || r3 | R4 )
        // (       T      ||  m |  B )
        tie(T, m, B) =
            continue_with<2>(R0, r1, R2, r3, R4);
    }
}
\end{lstlisting}
\end{tabular}
\end{center}

\caption{C++ implementation of the blocked right-looking algorithm of Figure~\ref{fig:LTLt_blk}. The use of ranges and range partitioning allows for strongly-typed identification of sub-matrices, vectors, and scalars. The tridiagonal factor $T$ is stored as an external vector \texttt{t} containing the sub-diagonal elements, while the triangular factor $L$ is stored in the lower part of $X$ with an implicit column shift. The diagonal elements of $L$ are stored as explicit ``1''s in order to enable combined updates such as the {\sc skew\_tridiag\_rankk} which involves $\lambda_{33}=1$.}
\label{fig:code}
\end{figure}

An important part of the FLAME workflow is that the representation in code of algorithms, themselves presented in FLAME notation, preserves correctness by having the code closely resemble the algorithm.
While it would be possible to extend the FLAMEC API~\cite{FLAME:API}, we chose to design and implement a new (prototype) C++ FLAME-like API.   

A representative implementation of the pivoted blocked right-looking algorithm is depicted in Fig.~\ref{fig:code}. 
We make use of MArray~\cite{marray}, which is a header-only DLA library developed by us, similar in spirit to Eigen \cite{eigen}, Armadillo \cite{armadillo}, Blitz++ \cite{veldhuizen_arrays_1998,veldhuizen_blitz_2000}, \textcolor{review2}{\textless{}T\textgreater{}LAPACK \cite{tlapack},} and others. 
We have extended this library with a range-based partitioning API which expresses the partition--repartition pattern of FLAME. 
Ranges are strongly typed, with singleton ranges (corresponding to partitions containing only one column or row) typed as scalar integers, and others typed as a C++ class representing the half-open range $[from,to)$. 
Indexing matrix or vector types with such ranges produces a result which depends on the range types, yielding a sub-matrix, sub-vector, or single scalar entry. Finally, BLAS and BLAS-like operations (besides simple scalar-vector operations which use the expression template functionality of MArray) use custom wrapper functions so that matrix lengths, strides, and other properties can be provided by the relevant objects. 
Since it is a prototype API, we don't give  details.

Importantly, even for unblocked algorithms, which are sensitive to overhead in terms of branches, function calls, and additional memory accesses (register spills, updating structs, etc.), typical compilers (in our case,  gcc 11.4.0) are capable of optimizing the combination of the while loop and range partition--repartition pattern into the same assembly as a plain for loop, with the exception of a small number of spurious conditional moves. As these instructions are spurious (referring to conditions already known), they can be perfectly predicted. This observation matches our profiling data which shows negligible overhead in the {\sc ltlt\_unb\_\emph{var}} function bodies.

\begin{wrapfigure}{r}{0.475\textwidth}

\begin{center}
\vspace{-12pt}
\includegraphics[width=0.475\textwidth]{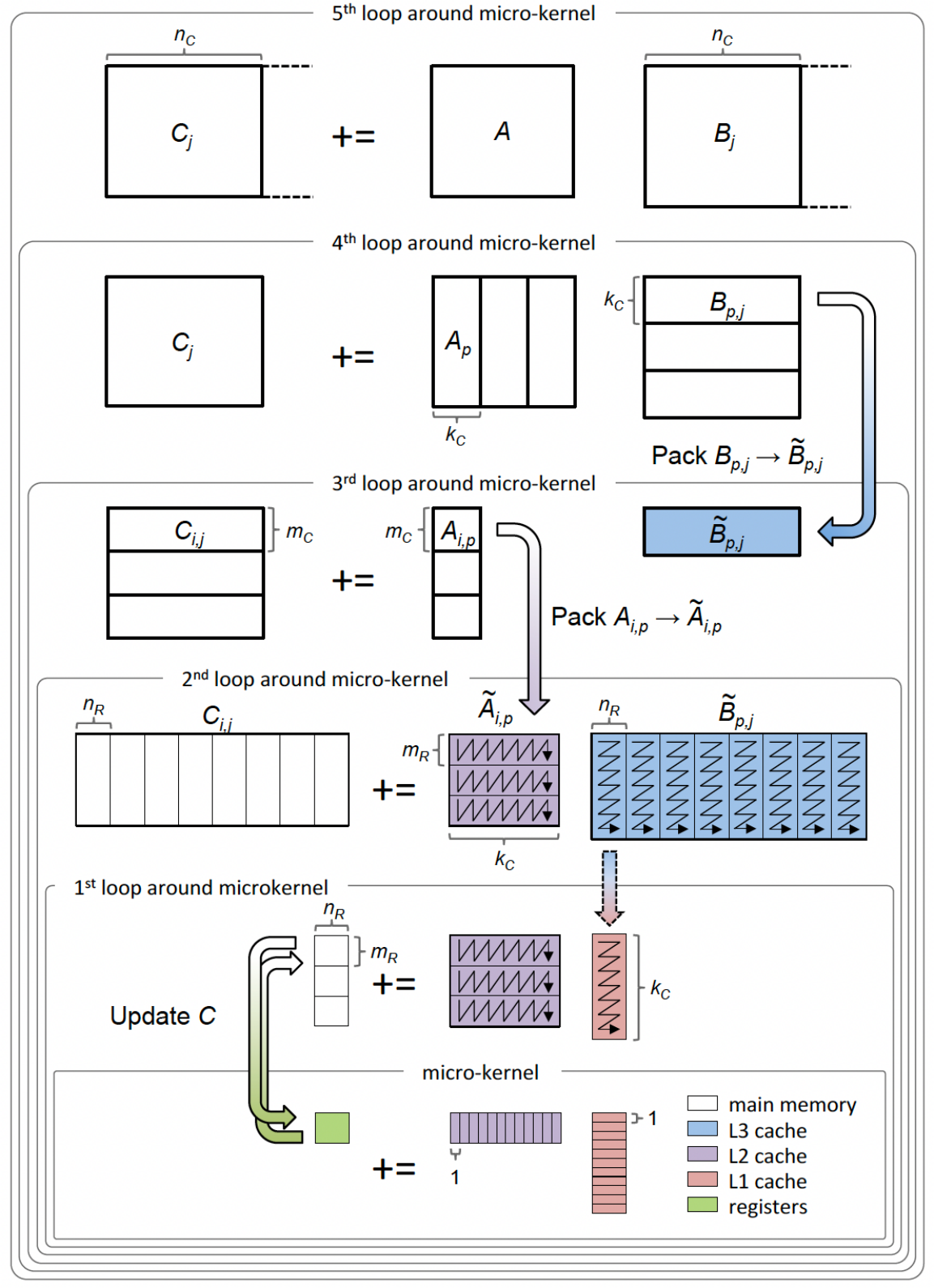}
\end{center}
\caption{The BLIS instantiation of Goto's algorithm~\cite{BLIS1,BLIS5}.}
\label{fig:BLIS}
\end{wrapfigure}

\subsection{Implementing level-3 BLAS-like operations}

The derived blocked algorithms cast most computations in terms of operations given in Fig.~\ref{fig:BLAS3-like}.
Take the ({\tt skew\_tridiag\_rankk}) $ C := \beta C + \alpha A T A^T $, where $ C $ is a skew-symmetric matrix, $ A $ is a general $ m \times k $, $ T $ is a skew-symmetric tridiagonal matrix, and only the lower-triangular part of $ C $ is updated .
This operation performs roughly $ m^2 k $ floating point operations (flops) on 
 $\mathcal{O}(mk)$ data.
 In order to cast this operation in terms of the traditional level-3 BLAS, 
 one could first compute 
 $ B = T A^T $ at a cost of $ \mathcal{O}( m k )$ computations and workspace and then update the lower triangular part of $ C := \beta C + \alpha A B $ with a call to {\sc gemmt}, which only affects half of $ X$.  While this does not increase the operation count significantly, it generates extra data movement, requires extra workspace, and makes code less readable.

 To overcome these limitations, one must understand the basics of how matrix-matrix multiplication (GEMM) is implemented on current CPUs via Goto's algorithm, depicted in Fig.~\ref{fig:BLIS}.
 Importantly, submatrices of $ A $ and $ B $ are packed at  strategic points in the algorithm in order to improve data locality.
 This exposes the opportunity for implementing GEMMT-SKTRI by  fusing the computation of  submatrices of $ B = T A^T $ with the packing of $ B $, thus avoiding the need for extra data movement and workspace.
 Similar techniques can be used to create implementations of BLAS-like operations on GPUs.

 With the advent of BLIS Release 2.0~\cite{BLIS20}, users can implement their own such ``sandwiched'' GEMM-like operations. This requires one to supply custom packing routines which incurs a  negligible increase in cost over standard packing because, although each element of $TA^T$ requires reading two elements of $L$ (due to the skew-symmetric tridiagonal structure), the accesses can take advantage of high temporal locality. The rest of the operation is handled without modification by BLIS, yielding the same overall performance as its standard level-3 operations\textcolor{review}{,
 while also} leveraging the infrastructure that already exists for GEMM within BLIS, including its multithreaded parallelization. 
 With this, we implemented the level-3 operations in Fig.~\ref{fig:BLAS3-like}.  

\subsection{Implementing level-2 BLAS-like operations}

While the leading-order cost of the blocked algorithms arises mostly from level-3  BLAS-like operations, the factorization of the current panel is typically performed with an unblocked algorithm and accounts for a nontrivial portion of the execution time. Thus, attention needs to be paid to how these are optimized. \textcolor{review}{In the absence of pivoting, recursive blocking (e.g. divide-and-conquer~\cite{5389353}) could be used, however as comparison to pivoted algorithms is a focus here, only one level of blocking is used in this work.}

The most costly updates in the various unblocked algorithms involve level-2 BLAS-like (matrix-vector) operations, which we tabulated in Fig.~\ref{fig:BLAS2-like}.
The\break{\sc skew\_rank2} (skew-symmetric rank-2 update) is a variation on the symmetric rank-2 update ({\sc syr2}) that is part the traditional BLAS and can be easily implemented via a minor modification of its implementation in, for example, BLIS.
The important thing is to implement this so that the triangular part of $ X $ is only read and written once.
The {\sc skew\_tridiag\_gemv} (matrix-vector multiplication with a vector that is first multiplied by a skew-symmetric tridiagonal vector) requires a cheap processing of the vector being multiplied.

In addition to having level-2 BLAS-like operations that capture the functionality needed, their effect on the overall performance also requires careful implementation.
Rather than casting the operations in terms of traditional BLAS, we wrote fused implementations using kernels from BLIS.
 Where possible, we applied loop unrolling and jamming (e.g., computing multiple {\sc axpy}s with a single loop) in order to improve register and prefetch pipeline usage.
 While parallelization of the new level-3 operations was inherited from BLIS's GEMM implementation, 
 at the time of this writing, BLIS did not parallelize level-2 functionality.
 For this  reason, we parallelized all needed level-2 operations (including the application of blocks of pivots used in  Fig.~\ref{fig:LTLt_blk}) with OpenMP~\cite{openmp}. 
 The benefits of these efforts are discussed next.

\section{Experimental results}
\label{sec:results}

To demonstrate the merits of the new algorithms and discussed optimizations, we performed a series of performance experiments.

\subsection{Experimental setup}  

The chosen system has  $2\times$ AMD EPYC 7763 processors (64 cores per socket) and 512 GiB of 8-channel (per socket) DDR4-3200 RAM. At the base frequency of 2.45 GHz, this provides 5.02 TFLOPs of peak double precision performance (2.51 TFLOPs per socket), although actual frequency depends on thermal load and available boost up to 3.5 GHz.


We compare our implementations to equivalent or computationally similar operations provided by several packages. We use standard LAPACK routines from Intel MKL 2022.2.1, OpenBLAS 0.3.26, and AMD AOCL 4.2. We also tested skew-symmetric $LTL^T$ factorizations with pivoting from PFAPACK and Pfaffine compiled from source using the same gcc 11.2.0 compiler as used for our code. A pre-release version of BLIS 2.0 was leveraged to implement the skew-symmetric and fused BLAS-like operations. We use OpenMP for parallelization, and pin threads to cores using the \texttt{OMP\_PLACES=cores} and \texttt{OMP\_PROC\_BIND=close} environment variables. In all cases we measure performance for double precision matrices, and the default \texttt{schedutil} frequency governor is used due to technical limitations.

\subsection{Impact of optimizations}\label{sec:opt}

\begin{figure*}[t]
    \centering
    \includegraphics[width=\textwidth]{images/opt_1_new.png}
    \vspace{-2em}
    \caption{Performance of blocked right-looking algorithms with the unblocked left-looking algorithm applied within blocks. Fusion and other optimizations are applied in steps (see text). \textcolor{review}{\textbf{Left column)}} without pivoting; \textcolor{review}{\textbf{right column)}} with pivoting. The algorithmic block size is \textbf{ab)} 128, \textbf{cd)} 192, \textbf{ef)} 256, and \textcolor{review}{\textbf{gh)}} 512. All calculations use 64 cores (1 socket). Notice the that the y-axis scaling is different for the graphs on the right.
    }
    \label{fig:opt}

\end{figure*}

The performance of the blocked right-looking factorization (with and without pivoting), using an unblocked left-looking algorithm for the panel factorization,  is presented in Fig.~\ref{fig:opt}, using one socket (64 threads). Performance for the implementation employing various steps of optimization (as discussed below) are presented.
Four different algorithmic block sizes are tested, from the top row to the bottom: 128, 192, 256, and 512. For reference: BLIS's {\sc gemm} is tuned to yield best performance for a rank-k update where $ k = 256 $ (parameter $ k_C $ in Fig.~\ref{fig:BLIS} for the processor used here). 
Using a larger  block size trades spending more time in the unblocked panel factorization with casting  more computation in each higher-performing level-3 update of the trailing matrix.  
As  observed, using a block size larger than 256 yields no further benefits since it breaks the update of the trailing matrix into multiple updates while casting more computation in the panel factorization.

In the same figure, we report the impact of the accumulation of a sequence steps that implement optimizations described in Section~\ref{sec:implementation}.
We start with:
\vspace{0.5em}
\begin{description}
    \item[Step~0] The blocked right-looking  algorithm (Variant 1) is employed, calling the unblocked left-looking algorithm.  Both $L$ and $T$ are stored in $X$ as they are computed and updates involving the diagonal of $L$ are unfused, as in (\ref{eqn:split_update1})--(\ref{eqn:split_update2}).
    The implementations call  BLIS's existing BLAS functionality plus the {\sc skew\_rank2} operation from Fig.~\ref{fig:BLAS2-like}, which is easily  derived from {\sc syr2}.
\end{description}
\vspace{0.5em}
Next, we apply optimizations that affect the performance of level-2 functionality:
\vspace{0.5em}
\begin{description}    
    \item[Step~1] Fused level-2 operations {\sc skew\_tridiag\_gemv} and {\sc gen\_rank2} are introduced, without parallelization.
    Optimizing the former has a very small (but positive) impact, as might be expected from the fusion of a vector computation with the {\sc gemv} operation.
    The latter reduces the number of memory accesses compared to two general rank-1 updates ({\sc ger}).

    \item[Step~2] All level-2 operations as well as block pivoting are parallelized.
This has a much larger effect, particularly at large core counts, as approximately 10 cores are required to fully saturate the memory bandwidth. This effect grows with algorithmic block size due to the increase in  work in level-2 operations.
\end{description}
\vspace{0.5em}
Finally, we apply optimizations that affect level-3 functionality:
\vspace{0.5em}
\begin{description}  
    \item[Step~3] 
    $T$ is stored in an external vector and fused updates are used which explicitly reference (unit) diagonal elements of $L$.
    This enables,  for the blocked right-looking  algorithm, the fusion of several matrix-vector and vector-vector operations with an ``adjacent'' {\sc skew\_tridiag\_rankk} in (\ref{eqn:split_update1})--(\ref{eqn:split_update2}).
    The effect on performance is small since the input sub-matrices for the matrix-vector updates come only from the current panel and hence they typically reside in L2 or L3 cache. The lack of fusion here does not increase the amount of data movement for the trailing block $X_{44}$, as is the case in Steps~4 and 5.
    \item[Step~4] The optimized fused level-3 operation {\sc skew\_tridiag\_rankk} is employed.
     This has a dramatic impact by reducing the number of times the trailing part of $ X  $ needs to be read from and written to memory.

    \item[Step~5] We switch to the fused blocked right-looking algorithm (Variant 2b),   reducing the number of times the trailing part of $ X  $ moves from and to memory.
    \NoShow{ of Fig.~\ref{fig:LTLt_blk}.}
\end{description}
\vspace{0.5em}
\NoShow{
{\color{red}\bf From arXiv paper:}

Successively
larger block sizes imply more time spent in the unblocked panel factorization, but also more aggregation into level-3 trailing
updates. Step 1 has a very small positive impact, as might be expected from the fusion of an essentially level-1 step into the
{\sc gemv} operation. Step 2 has a much larger effect, particularly at large core counts, as approximately 10 cores are required
to fully saturate the memory bandwidth. This effect grows with algorithmic block size due to the larger amount of work in
level-2 operations. The impact of Step 3 is also very small. At this step, essentially a {\sc skew\_tridiag\_gemv} is fused with an ``adjacent'' {\sc skew\_tridiag\_rankk} (at the blocked level, fusion at the unblocked level is less important). As the input sub-matrices for the {\sc skew\_tridiag\_gemv} come only from the current block they typically reside in L2 or L3 cache. The vector to update is typically in main
memory, but is only $\mathcal{O}(n)$ in size. This is in contrast to Steps 4 and 5, where now the operations which are fused operate
on $\mathcal{O}(n^2)$ data. The effect of these optimizations is much larger. The effect of fusing the trailing {\sc skew\_rank2} update (Step 5) is
particularly pronounced for small block sizes.
}
   Combined, these optimizations improve performance by more than $5\times$ for the unpivoted and $ 3 \times $ for the pivoted algorithms.

\subsection{Parallel scalability}

\begin{figure}[tb!]
    \centering
    \includegraphics[width=\textwidth]{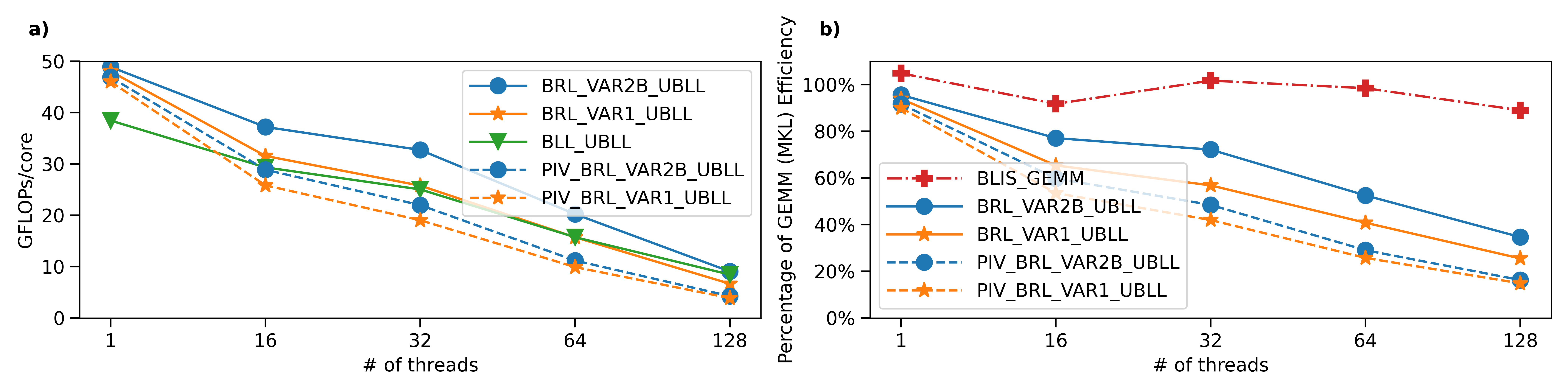}
    \vspace{-2em}
    \caption{Strong scaling   of  the various algorithm when  factoring a  $20,\!000\times 20,\!000$ matrix.  Left: GFLOPS/core.
    Right: Efficiency relative to that attained by Intel's MKL {\sc dgemm}.
    Since our implementations leverage BLIS, the performance of {\sc dgemm} in BLIS is also reported.
  }
    \label{fig:scaling}
\end{figure}

We measure the parallel scalability of our algorithms by performing a strong scaling experiment from one to all 128 cores of the system for a single $20,\!000\times 20,\!000$ skew-symmetric $LTL^T$ factorization. 
The results are presented in Fig.~\ref{fig:scaling}, with unpivoted and pivoted blocked algorithms represented by solid and dashed lines, respectively.

All algorithms involve non-trivial memory-bound work, in the panel factorization and pivoting. 
Additionally, the unfused blocked right-looking algorithm performs a level-2 {\sc skew\_rank2} operation which adds considerable overhead (see Section~\ref{sec:opt}). The impact of these operations can be clearly seen in the parallel scaling, as memory bandwidth only increases with core count up to $\sim 10$ cores, and latency is significantly impacted for memory traffic between chiplets (8-core chiplet dies or CCDs) and especially between sockets. The unpivoted algorithms experience a drop of $\sim 25\%$ in parallel efficiency (which is proportional to per-core performance) from 1 to 16 or 32 threads. Some of this is attributable to a drop in processor frequency due to thermal management. The single core results certainly show an effect of frequency boosting as the base clock single core peak is only 39.2 GFLOPs, while peak at max boost frequency is 56 GFLOPs. Assuming max boost, the fused right-looking algorithm (with or without pivoting) achieves $\sim 85\%$ of peak.

In order to measure this effect, we consider the MKL implementation of {\sc gemm} as the ``speed of light'' and present performance  of our algorithms relative to MKL's {\sc gemm} in Fig.~\ref{fig:scaling}\textbf{b}. Indeed, the relative performance for 16+ threads is higher than indicated by panel \textbf{a)}, although only slightly. The difference grows with core count, showing a progressive drop in processor frequency, although significant boosts are evident even for 32 or more cores. From panel \textbf{b)}, the biggest parallel efficiency drops are at 16 cores, likely due to exceeding a single 8-core cluster (CCD), and then at 64 threads and beyond. Based on profiling data, the main effect is due to increased time in level-2 operations, indicating main memory bandwidth as the limiting factor (inter-cache bandwidth would already be affected at core counts greater than 8 due to inter-CCD communication). In the two socket case, the impact of NUMA remote accesses is likely an additional factor. 

The scalability of the pivoted algorithms is even more severely affected. In particular, we measure the application of block pivots (see Fig.~\ref{fig:LTLt_blk} for example) as the largest overhead. This is primarily because the columns of $L$ to pivot are stored in column-major layout, while rows are swapped. Thus, accesses ``waste'' most of each cache line loaded and TLB reuse is poor, especially for very large matrices. We attempt to optimize stores by blocking across the row dimension to increase temporal locality. Overall, the best unpivoted algorithm achieves 41\% parallel efficiency (53\% of {\sc gemm}) at 64 cores, while the best pivoted algorithm achieves only 24\% (29\% of {\sc gemm}). We observe similar poor scalability of existing pivoted factorizations, e.g. pivoted Cholesky ({\sc pstrf}) and Bunch-Kaufman symmetric tridiagonalization ({\sc sytrf}) although data is not shown here.

\begin{figure*}[t]
    \centering
    
    \includegraphics[width=\textwidth]{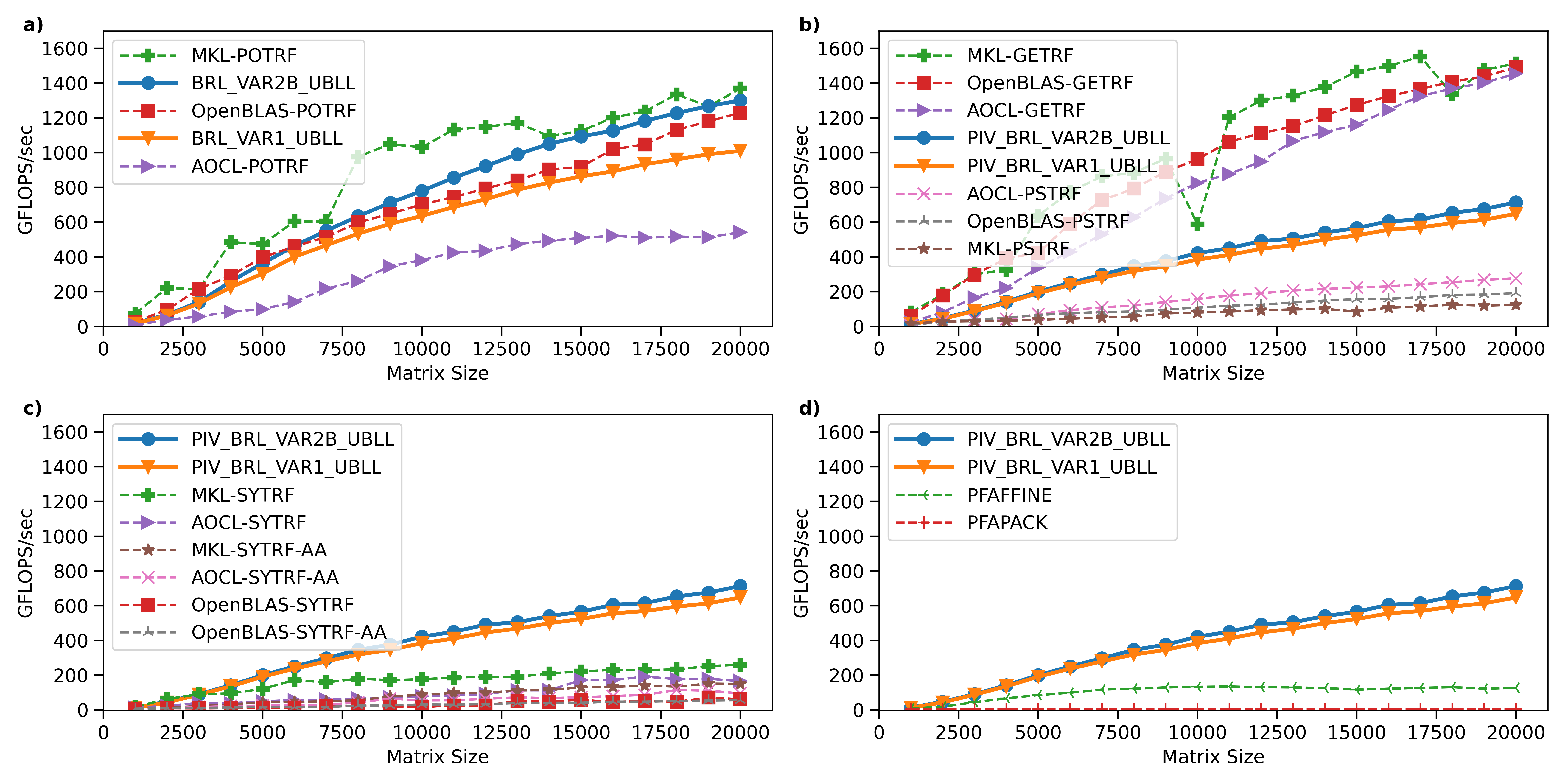}
    \vspace{-2em}
   \caption{Performance of our skew-symmetric $LTL^T$ implementations (labels as in Fig.~\ref{fig:scaling}) \textbf{a)} without pivoting compared to Cholesky factorization ({\sc potrf}), and with pivoting compared to \textbf{b)} \textcolor{review}{row}-pivoted $LU$ factorization ({\sc getrf}) and pivoted Cholesky ({\sc pstrf}), \textbf{c)} Bunch-Kaufman ({\sc sytrf}) and Rozlo\v{z}n\'{\i}k/Aasen ({\sc sytrf\_aa}) symmetric tridiagonalization, and \textbf{d)}  pivoted skew-symmetric $LTL^T$ factorization from PFAPACK and Pfaffine. All calculations use 64 cores (1 socket) and an optimal block size determined from experiments in Section~\ref{sec:opt}.}
    \label{fig:comparison}
\end{figure*}

\subsection{Comparison to similar operations}

 We now compare the performance of our implementation to those of related operations, performance optimization of which has been better studied.

\subsubsection{Cholesky factorization ({\sc potrf})}

 In Fig.~\ref{fig:comparison}\textbf{a}, we compare with various implementations of the Cholesky factorization, which performs similar computations and has (roughly) the same computational cost.  
 Since it does not require pivoting, we only compare with our unpivoted algorithms.
 Despite the fact that we do not employ multiple levels of blocking, 
 which is necessary to achieve the communication/IO lower bound \cite{ballard_cholesky} for many DLA algorithms, the performance is competitive.
 
 \NoShow{pivoted and pivoted operations which share computational and algorithmic similarities to $X=LTL^T$, but which are expected to be highly optimized given their ubiquity in scientific computing. Panel \textbf{a)} shows a comparison to Cholesky decomposition ({\sc dpotrf}) as implemented in .... Cholesky decomposition incurs roughly the same number of FLOPs as our algorithms, but is expected to be more efficient due to several factors which we do not take advantage of. The first is that while we limit ourselves to one level of blocking, this operation can employ multiple levels, each with a choice of algorithmic variant, which is necessary to achieve the communication/IO lower bound \cite{ballard_cholesky} for many DLA algorithms. The second is the use of level-3 BLAS operations ({\sc trsm} in the case of Cholesky) to update the trailing portion of blocks of the triangular factor. While the MKL implementation does exceed our performance comfortably, we find that the fused right-looking algorithm with left-looking unblocked updates is competitive with the OpenBLAS Cholesky implementation and much more efficient than that in AMD's AOCL. The unfused right-looking algorithm, as well as algorithms which use unblocked right-looking updates experience a significant drop in performance. This is almost exclusively due to the poor parallel performance of {\sc skr2}. We use a non-uniform thread partitioning scheme, but load imbalance and cache conflicts still significantly impact performance. The left-looking blocked algorithm, even though it does not perform an additional level-2 operation at the blocked level, falls behind the right-looking algorithm. As mentioned above, this is at least in part due to reduced parallelism in block updates, depending on the stage of the algorithm. 
 }

 \subsubsection{LU with partial pivoting and Cholesky with pivoting}
 In panel \textbf{b)}, we compare to 
 LU factorization with partial pivoting ({\sc getrf}) and Cholesky factorization with symmetric pivoting ({\sc pstrf}).
 
 LU factorization is typically among the most highly optimized operations due  to  its  importance.  It exhibits more opportunities for optimization since the pivoting is simpler,  multiple levels of blocking  can be employed, and the bulk of computation is in a general rank-k update rather than a variant on the symmetric rank-k update.  This explains the higher observed performance relative to our algorithms. \textcolor{review2}{Note that while LU factorization requires approximately twice the FLOPs as the other algorithms considered, this has been incorporated in the FLOP rate calculation.}

 The pivoted Cholesky 
  factorization ({\sc pstrf})
  has constraints that are more similar to the pivoted $ L T L^T $ factorization.  Performance of that operation is worse probably because not all optimization opportunities have been exhausted.

\subsubsection{Symmetric factorization of indefinite matrices}

Starting with a symmetric indefinite matrix $ X $, one can compute its  pivoted $ LDL^T $ factorization, where $ D $ has $ 1 \times 1 $ and $ 2 \times2$ blocks on the diagonal,
or a pivoted $ L TL^T $ factorization.
In the first case, a blocked  Bunch-Kaufman algorithm is available as  {\sc sytrf} in LAPACK and for the second case, some vendors have implemented the blocked variant of Rozlo\v{z}n\'{\i}k et al.~\cite{Miroslav2011}, implemented as {\sc sytrf\_aa}.
In Fig.~\ref{fig:comparison}\textbf{c} the performance of various   vendor implementations are compared with our implementations.
The better performance of our algorithms can largely be attributed to their use of BLAS-like operations that can be optimized (e.g., using the BLIS framework), but have instead been  cast in terms of traditional BLAS.

\subsection{Comparison to prior work}
Finally, in Fig.~\ref{fig:comparison}\textbf{d} we compare our algorithms to implementations of Wimmer's blocked right-looking algorithm from\break{}PFAPACK~\cite{pfapack} and Pfaffine~\cite{Pfaffine}. These implementations are very similar, with the latter switching from Fortran to C++ and replacing the reference, unoptimized {\sc skr2k} (= {\sc skew\_rank2k}) of PFAPACK with calls to an optimized version, implemented in terms of {\sc gemmt}. PFAPACK does not exceed 5.8 GFLOPs, compared to a peak of 135 GFLOPs for Pfaffian. Because we force the ``full'' factorization mode rather than the ``partial'' mode which is sufficient for computation of only the Pfaffian (as in Wimmer's algorithm), both implementations also perform twice as many FLOPs as our algorithms. For smaller matrices, the difference in FLOPs performed accounts for the observed performance gap, particularly for Pfaffine. However, for larger matrices the difference in performance becomes much larger than the expected factor of two.

\subsection{Discussion}

Overall, these results validate our new algorithms and the optimizations that we have applied, as the observed performance is roughly comparable to existing optimized implementations of related operations, and much higher than existing implementations of the  $LTL^T$ factorization (symmetric or skew-symmetric).

\section{Conclusion}

\label{sec:conclusion}

We have presented a number of  algorithms for computing the $ L T L^T $ factorization of a skew-symmetric matrix that result from applying the FLAME methodology.  These include classic algorithms like the Parlett-Reid and Aasen's algorithms (which  were proposed for symmetric $LTL^T$ 
but were modified for the skew-symmetric case) and Wimmer's algorithms.  It also yielded a number of new blocked algorithms that are straightforward to implement.  We exposed a  link between Wimmer's blocked algorithm and our novel fused blocked right-looking algorithms.  Together, these represent a coherent family of algorithms for this operation.

On the implementation side, we identified a number of level-2 and level-3 BLAS-like operations that support the encountered algorithms and discussed their high-performance implementation with BLIS.  A prototype C++ API for representing the algorithms in code was proposed and used to implement the various algorithms.  Performance experiments demonstrated the benefits of the new algorithms and high-performance kernels.

As the symmetric Partlett-Reid and Aasen algorithms have been modified for skew-symmetric matrices, our algorithms could also be modified back to the symmetric case. The improvement in performance in Figure~\ref{fig:comparison}\textbf{c} over existing symmetric triangular tridiagonalization implementations provides additional opportunities for the present work.

Finally, we note that the derivation of the presented algorithms has opened up new frontiers within the FLAME formalism. The formal inclusion of pivoting in linear algebra algorithm derivation is another exciting avenue of research. The advancement of these formal techniques, hand-in-hand with optimization strategies through fusion and the extension of traditional BLAS routines, presents a multitude of possibilities in the wider realm linear algebra as well as tensor operations.

\section*{Acknowledgments}
Computational resources for this research were provided by SMU’s O’Donnell Data Science and Research Computing Institute.

\bibliographystyle{siamplain}
\bibliography{biblio}
\end{document}